\documentclass[jou,10pt]{apa6}

\usepackage{multirow} 

\usepackage{amsmath}
\usepackage{amsfonts}
\usepackage{amssymb}
\usepackage{listings}
\makeatletter
\def\Let@{\def\\{\notag\math@cr}}
\makeatother
\usepackage{graphicx}
\usepackage[utf8]{inputenc}
\usepackage[T1]{fontenc}
\usepackage{float}

\usepackage[utf8]{inputenc}
\usepackage{amsmath,amssymb}
\newcommand{\textVerb}[1]{\texttt{\mbox{#1}}}
\usepackage{subcaption}
\usepackage[american]{babel}
\usepackage{csquotes}
\usepackage{apacite}
\usepackage{caption}
\usepackage{{url}}
\usepackage{rotating}
\usepackage{color}
\usepackage{array}

\usepackage{setspace}
\usepackage{booktabs}

\usepackage{etoolbox}
\BeforeBeginEnvironment{tabular}{\begin{singlespace}}
\AfterEndEnvironment{tabular}{\end{singlespace}}

\usepackage{rotating} 

\makeatletter
\newcommand*{\indep}{%
  \mathbin{%
    \mathpalette{\@indep}{}%
  }%
}
\newcommand*{\nindep}{%
  \mathbin{
    \mathpalette{\@indep}{\not}
  }%
}
\newcommand*{\@indep}[2]{%
  \sbox0{$#1\perp\m@th$}
  \sbox2{$#1=$}
  \sbox4{$#1\vcenter{}$}
  \rlap{\copy0}
  \dimen@=\dimexpr\ht2-\ht4-.2pt\relax
  \kern\dimen@
  {#2}%
  \kern\dimen@
  \copy0 
} 
\makeatother

\newcommand{\beginsupplement}{%
        \setcounter{table}{0}
        \renewcommand{\thetable}{S\arabic{table}}%
        \setcounter{figure}{0}
        \renewcommand{\thefigure}{S\arabic{figure}}%
     }
\renewcommand{\thefigure}{\arabic{figure}}

\usepackage{tocloft}
\makeatletter
\renewcommand{\@cftmaketoctitle}{}
\makeatother

\title{The Gaussian Graphical Model in Cross-sectional and Time-series Data}

\shorttitle{GAUSSIAN GRAPHICAL MODEL}

\author{ Sacha Epskamp,\textsuperscript{1} Lourens J.\ Waldorp,\textsuperscript{1} Ren\'{e} M\~{o}ttus,\textsuperscript{2} Denny Borsboom\textsuperscript{1}}

\twoaffiliations{ 1. University of Amsterdam: Department of Psychological Methods\\
	2. University of Edinburgh: Department of Psychology}

\abstract{

\tableofcontents

}

\begin{document}
\maketitle

\raggedbottom
\urlstyle{same}


\clearpage

\section{Abstract}

We discuss the Gaussian graphical model (GGM; an undirected network of partial correlation coefficients) and detail its utility as an exploratory data analysis tool. The GGM shows which variables predict one-another, allows for sparse modeling of covariance structures, and may highlight potential causal relationships between observed variables. We describe the utility in 3 kinds of psychological datasets: datasets in which consecutive cases are assumed independent (e.g., cross-sectional data), temporally ordered datasets (e.g., $n = 1$ time series), and a mixture of the 2 (e.g., $n > 1$ time series). In time-series analysis, the GGM can be used to model the residual structure of a vector-autoregression analysis (VAR), also termed \emph{graphical VAR}. Two network models can then be obtained: a temporal network and a contemporaneous network. When analyzing data from multiple subjects, a GGM can also be formed on the covariance structure of stationary means---the between-subjects network. We discuss the interpretation of these models and propose estimation methods to obtain these networks, which we implement in the R packages \emph{graphicalVAR} and \emph{mlVAR}. The methods are showcased in two empirical examples, and simulation studies on these methods are included in the supplementary materials.

\section{Introduction}

	There has been a surge of network models being applied to psychological datasets in recent years. This is consistent with a general call to conceptualize observed psychological processes not merely as indicative of latent common causes but rather as emergent behavior of complex, dynamical systems in which psychological, biological, and sociological components directly affect each other \cite{borsboom2011small,cramer2012dimensions,cramer2010comorbidity, schmittmann2013,van2006dynamical}. Such relationships are typically not known, and probabilistic network models \cite{koller2009probabilistic} are used to explore potential dynamical relationships between observables \cite{netpsych,van2014new}.  In this paper we aim to provide a methodological introduction to a powerful probabilistic network model applicable in exploratory data analysis, the Gaussian graphical model (GGM), and to propose how it can be used and interpreted in the analysis of both cross-sectional and time-series data.

\paragraph{Two lines of network research in psychology} We can currently distinguish two distinct and mostly separate lines of research in which networks are utilized on psychological datasets: the modeling of cross-sectional data and the modeling of intensive repeated measures in relatively short time frames (e.g., several times per day during several weeks). In cross-sectional modeling, a model is applied to a dataset in which multiple subjects are measured only once. The most popularly used methods estimate undirected network models---so-called pairwise Markov random fields  \cite{netpsych, murphy2012machine}. When the data are continuous and assumed normally distributed, the GGM can be estimated. The GGM estimates a network of \emph{partial correlation coefficients}---the correlation between two variables after conditioning on all other variables in the dataset  \cite{bootnetpaper}. This model is applied extensively to psychological data (e.g., \citeNP{cramer2012dimensions, fried2016, isvoranu, kossakowski2015, mcnally2015mental, van2015association}).

Researchers can obtain time-series data by using the experience sampling method (ESM; \citeNP{myin2009experience}), in which subjects are asked several times per day to fill out a short questionnaire using a device or smartphone app. Also, time-series data can arise from diary studies (e.g., a questionnaire completed at the end of the day) or physiological measurements, among other methods. Often, repeated measures of one or multiple participants are modeled through the use of (multi\-level) vector autoregressive (VAR) models, which estimate how well each variable predicts the measured variables at the next time point \cite{borsboom2013network}. These models are increasingly popular in assessing intraindividual dynamical structures (e.g., \citeNP{bringmann2013network, bringmann2015revealing, wigman2015exploring}). 

Estimating the GGM is not limited to cross-sectional data; the model merely does not take temporal information into account. As such, the lines of research on network modeling of cross-sectional data and time-series data can naturally be combined. First, GGM models can readily be estimated on repeated measures, if these can be assumed to be temporally independent. Second, as the VAR model can be seen as a generalization of the GGM that takes violations of independence between consecutive cases into account; the GGM can be used to model the contemporaneous time level of a time-series analysis. Finally, the between-subjects effects of $n>1$ studies can also be modeled through the use of the GGM.

\paragraph{Outline}	We show that in time-series modeling the GGM allows researchers to extend the modeling framework to incorporate contemporaneous and between-subjects effects. We do this by building up the model complexity in three steps: (1) when cases can be assumed to be independent (e.g., cross-sectional data or repeated measures in which no auto-regression is assumed), (2) temporally ordered data (e.g., $n=1$ time-series data or $n>1$ time-series data where no individual differences are assumed), and (3) temporally ordered data from multiple subjects (e.g., $n>1$ time series). The final level of model complexity leads to a novel contribution of this paper: separation of variance into \emph{contemporaneous}, \emph{temporal}, and \emph{between-subjects} network structures. We propose novel estimation procedures to estimate these models, which we have implemented in two free software packages: \emph{mlVAR},\footnote{CRAN link: \url{http://cran.r-project.org/package=mlVAR}\\Github link (developmental): \url{http://www.github.com/SachaEpskamp/mlVAR}.} and \emph{graphicalVAR}.\footnote{CRAN link: \url{http://cran.r-project.org/package=graphicalVAR}\\Github link (developmental): \url{http://www.github.com/SachaEpskamp/graphicalVAR}.} We furthermore expand on existing literature by providing a comprehensive methodological discussion of the GGM, by comparing the GGM to structural equation modeling (SEM; \citeNP{kaplan2008structural, wright1921correlation}), by providing overviews of estimation methods and software packages useable in each kind of dataset and by discussing the interpretation of networks estimated at the contemporaneous and between-subjects levels. We showcase network models estimated from $n>1$ time-series data in two empirical examples by reanalyzing existing datasets \cite{bringmann2013network, geschwind2011mindfulness, mottus2016within}. In the supplementary materials, we provide codes to perform the analyses and we assess the performance of these methods in large-scale simulation studies. To aid the reader in the various different terms used in this paper, we have included a glossary of terms in Appendix~A.

\section{The Gaussian Graphical Model}

	Let $\pmb{y}_C^\top = \begin{bmatrix} Y_{C1} & Y_{C2} & \ldots & Y_{Cm} \end{bmatrix}$ denote a random vector with $\pmb{y}_c$ as its realization.\footnote{We use capitalized subscripts to denote random variables and lower case subscripts to denote fixed variables. A variable can potentially be fixed with respect to one subscript but random with respect to another. Supplementary materials section 1 contains a complete overview of the notation used in this paper.}  We assume $\pmb{y}_C$ is centered\footnote{Because we assume data to be centered, we do not need to model the (grand) mean vector. This simplifies notation.} and normally distributed with some variance--covariance matrix $\pmb{\Sigma}$:
\begin{equation}
\label{yisnormal}
\pmb{y}_C \sim N( \pmb{0}, \pmb{\Sigma}).
\end{equation}
The subscript $C$ denotes a case (a row in the spreadsheet). We currently do not define the nature of the observed variables. Thus, $\pmb{y}_C$ can consist of variables relating to one or more subjects, could contain repeated measures on one or more variables, could contain variables of a single subject that do not vary within-subject, and so forth. Consider three examples: (1) $Y_1$ could represent the level of anxiety of subject $p$ on day 1 and $Y_2$ the level of anxiety of subject $p$ on day 2, (2) $Y_1$ could represent the length of subject $p$ and $Y_2$ the number of times subject $p$ bumps his or her head, and (3) $Y_1$ could represent the number of cigarettes subject $p$ smokes per day and $Y_2$ the number of cigarettes another subject $p+1$ smokes per day (case $C$ then represents a dyadic pair). 

\paragraph{Partial correlation networks} Assuming multivariate normality, $\pmb{\Sigma}$ encodes all the information necessary to determine how the observed measures relate to one another. However, we will not focus on $\pmb{\Sigma}$ in this paper but rather on its inverse---the \emph{precision matrix} $\pmb{K}$:
\begin{equation*}
\pmb{K} = \pmb{\Sigma}^{-1}.
\end{equation*}
Of particular importance is that the precision matrix can be standardized to encode partial correlation coefficients of two variables, given all other variables (dropping subscript $C$ for notational clarity; \citeNP{lauritzen1996graphical}):\footnote{This relationship can be traced back much further. For example, \citeA{willemTalk} traced this relationship back to the work of \citeA{guttman1938note}.}
\begin{equation}
\label{eq:parcor}
\mathrm{Cor}\left(Y_i,Y_j \mid \pmb{y}_{-(i,j)}\right) = - \frac{\kappa_{ij}}{\sqrt{\kappa_{ii}} \sqrt{\kappa_{jj}}},
\end{equation}
in which $\kappa_{ij}$ denotes an element of $\pmb{K}$, and $\pmb{y}_{-(i,j)}$ denotes the set of variables without $i$ and $j$. These partial correlations can be graphically displayed in a weighted network, in which each variable $Y_i$ is represented as a node, and connections (edges) between these nodes represent the partial correlation between two variables. When the partial correlation (thus the corresponding element in $\pmb{K}$) equals zero, no edge is drawn. Thus, modeling the inverse variance--covariance matrix, such that every nonzero element is treated as a freely estimated parameter, allows for a sparse model for $\pmb{\Sigma}$ (i.e., every element in $\pmb{\Sigma}$ may be nonzero while some elements in $\pmb{K}$ are zero; \citeNP{epskampPsychometrika}). Such a model is termed a GGM (\citeNP{lauritzen1996graphical}). Of note, when the sample-variance--covariance matrix is inverted and standardized, no partial correlation will be exactly equal to zero and the GGM will therefore be saturated. To obtain a sparse model with testable implications, in this paper partial correlations are forced to zero either by using thresholding rules or regularization techniques.

When drawing a GGM as a network (often termed a \emph{partial correlation network}), positive partial correlations are typically visualized with blue or green edges and negative partial correlations with red edges,\footnote{Many publications make use of the default color setup used in \emph{qgraph} \cite{jssv048i04}: green for positive edges and red for negative edges. A later version of \emph{qgraph} includes the option \textVerb{theme = "colorblind"} using a more colorblind-friendly coloring scheme and setting the positive edge color to blue. This option has been used for all graphs in this paper. Note that some publications (e.g., \citeNP{noemi5}) also use blue and red edges but use red to denote positive and blue to denote negative effects akin to a heat map.} and the absolute strength of a partial correlation is represented by the width and saturation of an edge \cite{jssv048i04}. When a partial correlation is zero, we draw no edge between two nodes. As such, the GGM can be seen as a network model of conditional associations; no edge indicates that two variables are independent after conditioning on all other variables in the dataset. This allows us to model conditional associations, which we might expect to be zero, rather than marginal associations, which we rarely expect to be zero \cite{meehl1990summaries}.

\begin{figure}
\centering
\includegraphics[width=1\linewidth]{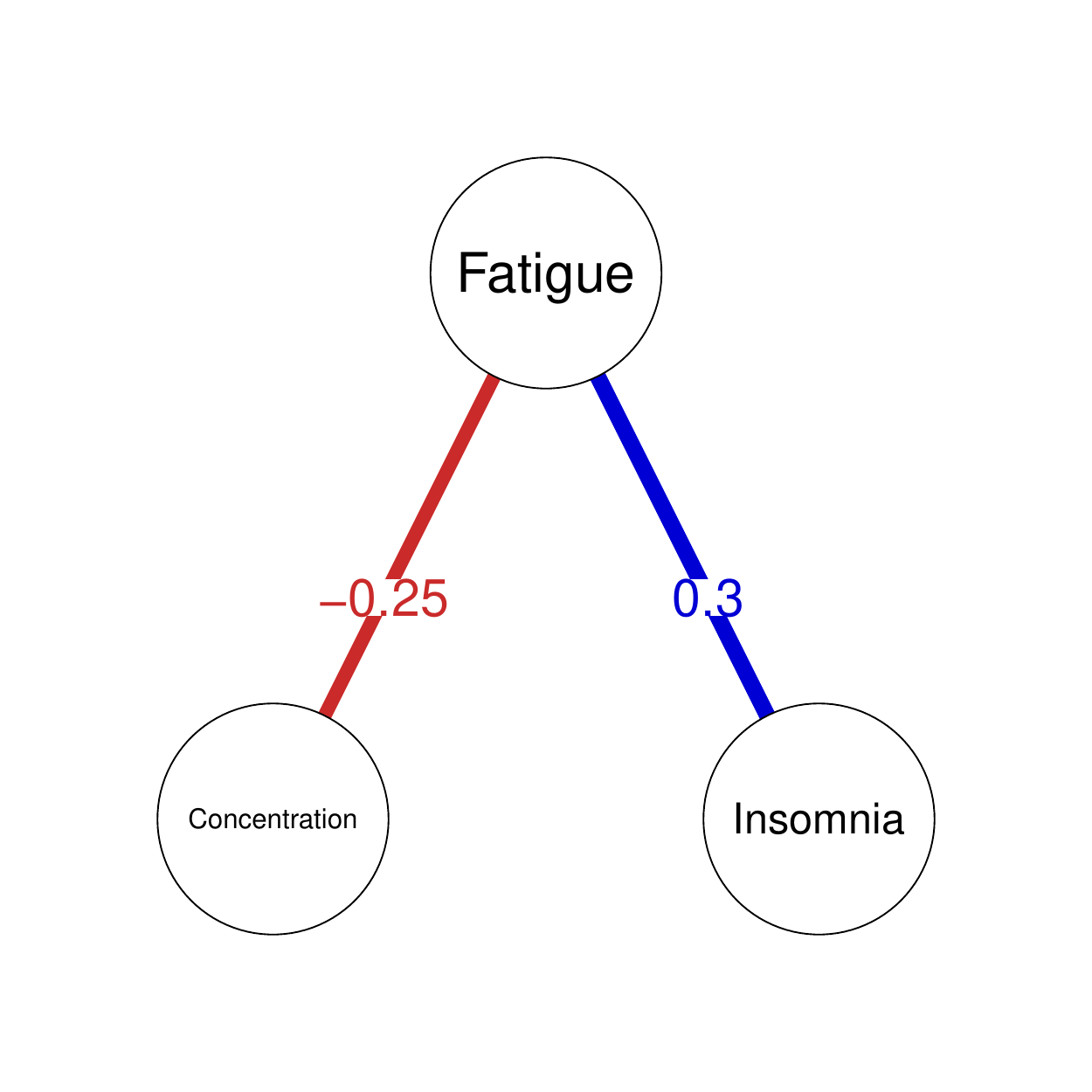}
\caption{A hypothetical example of a GGM on psychological variables. Nodes represent someone's ability to concentrate, someone's level of fatigue, and someone's level of insomnia. Connections between the nodes, termed \emph{edges}, represent partial correlation coefficients between two variables after conditioning on the third. Blue edges indicate positive partial correlations, red edges indicate negative partial correlations, and the width and saturation of an edge corresponds to the absolute value of the partial correlation. }
\label{dynamics:fig:GGM}
\end{figure}

To exemplify the above, suppose for three variables ``fatigue'', ``concentration problems'' and ``insomnia'' the true variance--covariance matrix is:
\begin{equation*}
\pmb{\Sigma} = \begin{bmatrix}
1 & -0.26 & 0.31 \\
-0.26 & 1 & -0.08 \\
0.31 & -0.08 & 1
\end{bmatrix}.
\end{equation*}
To model this matrix, we need six parameters (three covariances and three variances). The corresponding true precision matrix becomes:
\begin{equation*}
\pmb{K} = \pmb{\Sigma}^{-1} = \begin{bmatrix}
1.18 & 0.28 & -0.34 \\
0.28 & 1.07 & 0 \\
-0.34 & 0 & 1.11
\end{bmatrix}.
\end{equation*}
Similar to SEM, a model can be devised that perfectly explains this pattern using only five parameters, because one of the elements in $\pmb{K}$  can be constrained to be zero \cite{epskampPsychometrika}. We can now standardize this matrix and make the off-diagonal elements negative (Equation~\eqref{eq:parcor}) to obtain the partial correlation matrix, which we will denote $\pmb{R}$:
\begin{equation*}
\pmb{R} = \begin{bmatrix}
1 & -0.25 & 0.3 \\
-0.25 & 1 & 0 \\
0.3 & 0 & 1
\end{bmatrix}.
\end{equation*}
This matrix can be used to draw a network as is shown in Figure~\ref{dynamics:fig:GGM}. This figure shows that someone who is tired is also more likely to suffer from concentration problems and insomnia. Furthermore, this network shows that the correlation between insomnia and concentration problems can be explained by the relationships of both variables with fatigue: concentration problems and insomnia are conditionally independent given the level of fatigue.

\paragraph{Interpreting GGMs}  This paper concerns the exploratory estimation of GGMs  from various sources of data, without prior knowledge on the model structure. Such undirected network models can be interpreted in strikingly different ways, ranging from a no causal interpretation to a strong causal interpretation:
\begin{enumerate}
\item  \emph{Predictive effects}. The GGM can be interpreted without any causal interpretation and used merely as a tool to show which variables predict one-another. Interpreting the parameters associated with the model $A \rightarrow B \rightarrow C$ requires a causal interpretation, while the predictive quality between these nodes can directly be obtained from the equivalent GGM $A$ --- $B$ --- $C$: only information on node $B$ is needed when predicting $A$ or $C$. As such, the GGM can always be interpreted to show predictive effects and offers a powerful exploratory tool to map out multicollinearity.
\item  \emph{Indicative of causal effects}. The GGM is closely tied to causal modeling. If a causal model between observed variables generated the data, then an edge $A$ -- $B$ appears in the GGM only if there is a causal link between the variables (e.g., $A \rightarrow B$ or $A \leftarrow B$), or if both variables cause a third variable in the data (e.g., $A \rightarrow C \leftarrow B$). Exploratory estimation of such models relies on stringent assumptions (e.g., acyclicity), suffers from a problem of many equivalent models, and may lead to over-saturated models.  The GGM, on the other hand, is well identified and does not feature equivalent models. Therefore, at the cost of losing information on the direction of effect, exploratory search algorithms perform well in identifying a GGM. Because of this close tie to causal modeling, edges in the GGM may be interpreted as indicative of potential causal pathways.
\item  \emph{Causal generating model}. Undirected network models have a long history of being used as data generating models in diverse scientific fields such as statistical physics \cite{murphy2012machine}. For example, in a simple ferromagnetic Ising model of two particles that tend to be aligned \cite{netpsych}, $A$ --- $B$, intervening on $A$ would impact $B$ and intervening on $B$ would impact $A$. To this end, undirected network models allow for a unique causal interpretation: one of genuine symmetric effects. This interpretation is discussed often in the literature on network psychometrics, and used in complexity research demonstrating emergent phenomena (e.g., the positive manifold or phase transitions) that may occur in such a network of cellular automata \cite{van2006dynamical,cramer2016major,dalege2016toward,kruis2}. 
\end{enumerate}
In addition, the GGM is closely tied to factor analysis, allowing for extensions to factor modeling through the use of network modeling \cite{epskampPsychometrika}. The main focus of this paper is discussing the second interpretation, while also describing how the GGM may be used to show predictive effects. We detail these first two points below by first showing how partial correlation coefficients correspond to multiple regression coefficients and next discussing the relationships between the GGM and SEM. Point 3 follows from observing that the GGM is directly related to similar undirected models such as the Ising model \cite{ising1925beitrag}. A discussion on the causal interpretation of such models is beyond the scope of this paper, and we refer the reader for this topic to \citeA{netpsych} and \citeA{van2014new}.


\subsection{The Gaussian Graphical Model and Multiple Regressions}

An edge in a GGM indicates that one node predicts a connected node after controlling for all other nodes in the network. This can also be shown in the relationship between coefficients obtained from least-squares prediction and the inverse variance--covariance matrix. Let $\pmb{\Gamma}$ represent an $k \times k$ matrix with zeros on the diagonal. Furthermore, let $\pmb{\gamma}_{i,-(i)}$ represent the $i$-th row of $\pmb{\Gamma}$  without the $i$-th element (as the diagonal is set to zero), which contains the regression coefficients obtained in a multiple regression model:
\begin{equation*}
\label{dynamics:eq:nodewiseReg}
y_{ci} = \tau + \pmb{\gamma}_{i,-(i)} \pmb{y}_{c,-(i)} + \varepsilon_{ci}.
\end{equation*}
As such, $\gamma_{ij}$ encodes how well the $j$th variable predicts the $i$th variable. This predictive effect is naturally symmetric; if knowing someone's level of insomnia predicts his or her level of fatigue, then conversely knowing someone's level of fatigue allows us to predict his or her level of insomnia. As a result, $\gamma_{ij}$ is proportional to $\gamma_{ji}$. There is a direct relationship between these regression coefficients and the inverse variance--covariance matrix \cite{meinshausen2006high}. Let $\pmb{D}$ denote a diagonal matrix on which the $i$th diagonal element is the inverse of the $i$th residual variance: $d_{ii} = 1/\mathrm{Var}(\varepsilon_{Ci})$. As a result, it can be shown \cite{pourahmadi2011covariance} that\footnote{This expression may differ by a scalar, depending on the estimation method. For example, by default R computes the variance--covariance matrix by using $n-1$ in the denominator, but computes $\mathrm{Var}(\varepsilon_{Ci})$ by using $n - m$ in the denominator. This denominator is cancelled out in Equation~\eqref{eq:parcor} when standardizing to partial correlation coefficients.}
\begin{equation}
\label{dynamics:eq:regToGGM}
\pmb{K} = \pmb{D} \left(\pmb{I} - \pmb{\Gamma}\right).
\end{equation}
Thus, $\kappa_{ij}$ is proportional to both $\gamma_{ij}$ and  $\gamma_{ji}$; a zero in the inverse variance--covariance matrix indicates that one variable does not predict another. Consequently, the network tells us something about the extent to which variables predict each other. This predictive quality is the cornerstone for how such network models are often applied \cite{hastie2015statistical}, for example in recommender-systems that recommend users on products they might like depending on which products the user already liked \cite{marsman2017note}. In addition to these applications and aiding the interpretation of GGM models, this relationship between multiple regression and undirected network edges plays a crucial role in many network estimation procedures \cite{meinshausen2006high,van2014new,jonas2}, including the methods discussed below in this paper.

\subsection{The Gaussian Graphical Model and Structural Equation Modeling}

Let $\pmb{\eta}_C$ represent a set of unobserved variables, which we assume to be jointly normally distributed with $\pmb{y}_C$. Then, we can form an encompassing framework for several possible generating models:\footnote{This expression should not be confused with Equation~(3), in which $\pmb{\Gamma}$ is obtained by performing univariate multiple regressions in which error terms are not independent.}
\begin{align}
    \pmb{y}_c &= \pmb{B} \pmb{y}_c + \pmb{\Lambda} \pmb{\eta}_c + \pmb{\varepsilon}_c  \label{generating}\\
    \pmb{\varepsilon}_C &\sim N(\pmb{0}, \pmb{\Phi}) \\
     \pmb{\eta}_C &\sim N(\pmb{0}, \pmb{\Psi}),
\end{align}
in which $\pmb{\Phi}$ is a diagonal matrix\footnote{This matrix is often denoted using the greek letter $\pmb{\Theta}$ instead of $\pmb{\Phi}$. We use $\pmb{\Phi}$ here to avoid confusion with the contemporaneous variance--covariance matrix used below, which is not diagonal.}, indicating that after conditioning on all causes the variables are independent, $\pmb{B}$ is a square matrix with zeros on the diagonal of causal effects between observed variables, and $\pmb{\Lambda}$ is a factor-loading matrix. The variance--covariance matrix of $\pmb{\eta}_C$ may in turn be modeled in various ways to achieve complicated model setups. The expression above is well-known in SEM, which allows for confirmatory testing of causal models. In \emph{exploratory} estimation, one could assume no latent variables exist and aim estimate $\pmb{B}$ (causal models), or one could assume no relationships between observed variables exist and aim to estimate $\pmb{\Lambda}$ (factor models). We contrast both to the GGM below.

\subsubsection{Causal models} 

Suppose there are no unobserved causes to any of the variables in $\pmb{y}_C$, and the variables in $\pmb{y}_C$ are only caused by other variables in $\pmb{y}_C$. The corresponding model for $\pmb{\Sigma}$ becomes:
\begin{equation}
\pmb{\Sigma} = \left( \pmb{I} - \pmb{B}\right)^{-1} \pmb{\Phi}\left( \pmb{I} - \pmb{B}\right)^{-1\top}. \label{CausalModel}
\end{equation}
In this expression, $\pmb{B}$ can now be seen to encode the causal model \cite{pearl2000causality}. Table~\ref{causaltable} summarizes the comparison between such causal models and GGMs. Although useful for generating data and confirmatory testing, we can see two problems in exploratory estimation of $\pmb{B}$ without any prior knowledge. First, if $m$ variables are included, $\pmb{\Sigma}$ contains $m(m+1)/2$ elements, while $\pmb{\Phi}$ contains $m$ parameters and $\pmb{B}$ contains $m(m-1)$ parameters. As a result, the model above is underidentified without stringent restrictions on $\pmb{B}$. One assumption is that $\pmb{y}_C$ can be ordered such that $\pmb{B}$ is lower triangular, indicating that if this matrix is used to draw a directed graph---a graph in which $A \rightarrow B$ indicates that $A$ causes $B$---that graph does not contain any cycles, meaning that  directed edges cannot be traced from any node back to itself (e.g., $A \rightarrow B \rightarrow A$). Such a graph is called a directed acyclic graph (DAG; \citeNP{kalisch2007estimating,pearl2000causality}). If repeated measures are available at the correct time scale, reciprocal effects and cycles can often be adequately modeled as acyclic effects unfolding over time. Without such information, cycles can be modeled and can be identified when exogenous variables are present (such as the weather, time, or, depending on the modeling framework, lagged variables; \citeNP{rigdon1995necessary}), but the interpretation of such cycles is not without problems \cite{hayduk2009finite}. Several software packages exist that aim to find such a DAG (e.g., \emph{pcalg}, \citeNP{kalisch2012causal}; \emph{bnlearn}, \citeNP{Scutari2010}). The assumption of acyclicity, however, is debatable in the context of psychological variables \cite{schmittmann2013} because many effects can be plausibly assumed cyclic (e.g., fatigue $\rightarrow$ concentration problems $\rightarrow$ stress $\rightarrow$ fatigue).

\begin{table*}
	\centering
	{\footnotesize
	\caption{Overview of Causal Models (Directed Networks) and Gaussian Graphical Models (Undirected Networks)	\label{causaltable}}	
		\begin{tabular}{l p{5cm} p{5cm}}
			\toprule
			& Causal model & Gaussian graphical model  \\
			\midrule 
			$\pmb{\Sigma}^{-1} =$ & $\left( \pmb{I} - \pmb{B}\right)^{\top} \pmb{\Phi}^{-1}\left( \pmb{I} - \pmb{B}\right)\quad =$ & $\pmb{K}$ \\[0.5cm]
			$A \!\perp\!\!\!\perp C \mid B$ & \includegraphics[width=5cm,page=1]{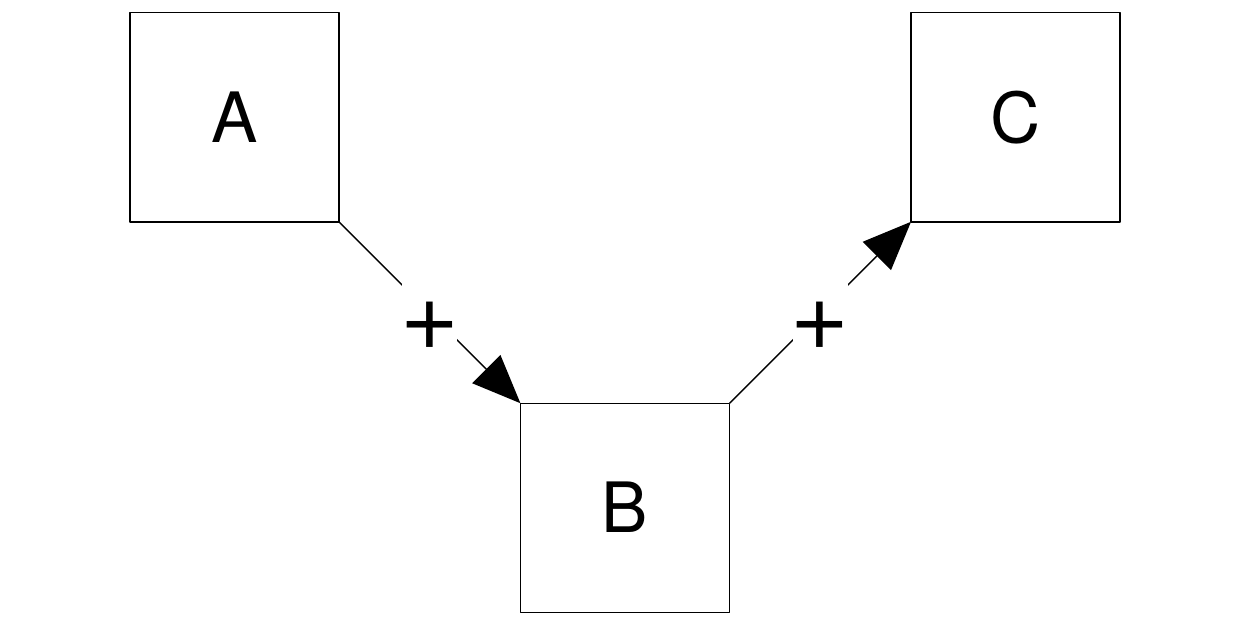}&
			\includegraphics[width=5cm,page=4]{Table1.pdf}\\
			$A \!\perp\!\!\!\perp C \mid B$ & \includegraphics[width=5cm,page=2]{Table1.pdf}&
			\includegraphics[width=5cm,page=4]{Table1.pdf}\\
			$A \not\!\perp\!\!\!\perp C \mid B$ & \includegraphics[width=5cm,page=3]{Table1.pdf}&
			\includegraphics[width=5cm,page=5]{Table1.pdf}\\[0.5cm]
			R packages (confirmatory) & Any SEM package & \emph{lvnet} (fit measures); \emph{qgraph} (fit measures); \emph{ggm} (estimation only); \emph{GLASSO} (estimation only) \\[0.5cm]
			R packages (exploratory) & \emph{pcalg}; \emph{bnlearn} & \emph{qgraph} (EBICglasso function); \emph{GLASSO} (no automatic tuning parameter selection); \emph{huge}; \emph{parcor}; \emph{BDgraph}; \emph{lvnet} (for GGM at latent or residual level of SEM) \\[0.5cm]
			Pros &  Causal interpretation; allows for confirmatory testing of causal hypotheses; can detect common effect structures &  No equivalent models; fast structure and parameter estimation using LASSO; edges parametrizable as partial correlation coefficients; edges interpretable as predictive effects; latent variables result in clusters; edges can be indicative of potential causal effects \\[0.5cm]
			Cons & Exploratory estimation requires assumption of acyclicity; many equivalent models; direction of effect poorly or not identified; strongly depends on assumption of no latent variables  &  No direction of effect; common effect structure can induce spurious edge; LASSO estimation assumes true model is sparse  \\
			\bottomrule
		\end{tabular}
	}
\end{table*}

Second, the same structure for $\pmb{\Sigma}$ can be obtained under various different specifications of $\pmb{B}$. Thus, many equivalent models can lead to exactly the same fit. This can be seen because several matrix decompositions of $\pmb{\Sigma}$, such as a Cholesky decomposition or an eigendecomposition, can be used to produce equivalently fitting $\pmb{B}$. The problem of equivalent models is also well-known in the literature on directed networks and SEM \cite{maccallum1993problem,pearl2000causality}. For example, the following three causal models are not statistically distinguishable:
\begin{enumerate}
\item Concentration $\rightarrow$ Fatigue $\rightarrow$ Insomnia
\item Concentration $\leftarrow$ Fatigue $\rightarrow$ Insomnia
\item Concentration $\leftarrow$ Fatigue $\leftarrow$ Insomnia
\end{enumerate}
All three models only imply that concentration and insomnia are conditionally independent given fatigue. With more variables, the number of potential equivalent models increases drastically, making it evident that model search is likely to fail. At best, exploratory estimation can result in a set of equally plausible DAGs (an \emph{equivalence class}; \citeNP{drton2017structure}), each differently parameterized and each leading to different strong causal hypotheses.

\paragraph{Causal modeling and the GGM} The undirected GGM offers an attractive alternative to exploratory DAG estimation: the GGM is saturated rather than overidentified if all edges are present ($\pmb{K}$ contains the same number of unique elements as $\pmb{\Sigma}$), does not feature equivalent models\footnote{Note that the uniqueness of the GGM relates to the psychometric model: for every $\pmb{\Sigma}$ there is only one unique inverse $\pmb{K}$ and vice versa. When estimating a GGM from data, different estimation methods may lead to different estimated GGMs} (there is only one unique inverse for $\pmb{\Sigma}$), does not suffer from a questionable direction of causal effect, does not require the assumption of acyclicity, and is easily parameterized using partial correlation coefficients \cite{epskampPsychometrika}. These benefits come, however, at the cost of losing information on the direction of effect. To investigate the structure of a GGM under the causal model of Equation~\eqref{CausalModel}, in which observed variables can only be caused by other observed variables, we can invert that expression to obtain:
\begin{equation}
\pmb{K} = \left( \pmb{I} - \pmb{B}\right)^{\top} \pmb{\Phi}^{-1}\left( \pmb{I} - \pmb{B}\right), \label{GGMtoDAG}
\end{equation}
in which $\pmb{\Phi}^{-1}$ is still a diagonal matrix. It becomes evident that there is no longer a matrix inversion needed and that the sparsity in $\pmb{B}$ directly corresponds to the sparsity in $\pmb{K}$; the GGM thus acts on the same level as causal modeling. We can derive that $\kappa_{ij}$ equals zero if there is no directed edge between node $i$ and $j$ (e.g., $Y_i \rightarrow Y_j$ or $Y_i \leftarrow Y_j$) and if there is no common effect of node $i$ and node $j$ (e.g., $Y_i \rightarrow Y_k \leftarrow Y_j$; \citeNP{koller2009probabilistic})\footnote{A common effect node is also termed a ``collider'' in the literature on causal modeling.}. Thus, assuming a causal model as in Equation~\eqref{CausalModel} generated the data, an edge in a GGM emerges as a result of a direct causal effect between the variables, or as a result of the fact that both variables have a common effect on a third variable. Note that within the causal model of Equation~\eqref{CausalModel}, there are no latent common causes by assumption. Edges in the GGM can therefore be indicative of potential causal effects. 

\paragraph{A note on common effects} As mentioned above and shown in Table~\ref{causaltable}, conditioning on a common effect may induce a spurious edge in the GGM. In this case, the sign of the edge can be informative: two positive causal effects from two variables on a third lead to a negative partial correlation. As such, when observing an edge of an unexpected sign in the GGM, this may be indicative of a common effect, especially when the marginal correlation coefficient between the two variables was of the different sign. It should also be noted that conditioning on a common effect might cancel out a weak effect between two variables. In addition, because edges may be induced due to conditioning on a common effect, the GGM does not estimate a \emph{skeleton graph}, a causal network with arrowheads removed (an edge may be in the GGM that is not in the causal model). Skeleton graphs can also be estimated from data (e.g., \citeNP{kalisch2010}), but are not parameterized and rely on many separate conditional independence tests, potentially leading to power issues.

\subsubsection{Factor models}

Suppose that, instead of assuming no unobserved causes as in \eqref{CausalModel}, we take the generating model of \eqref{generating} and \emph{only} allow for unobserved causes of the observed variables. Then, \eqref{generating} reduces to the well known factor model \cite{brown2014confirmatory}. The corresponding model for $\pmb{\Sigma}$ now becomes:
\[
\pmb{\Sigma} = \pmb{\Lambda} \pmb{\Psi} \pmb{\Lambda}^{\top} + \pmb{\Phi},
\]
which can subsequently be inverted to obtain an expression for the equivalent GGM. \citeA{EGA} provide a detailed derivation of this inverted expression and show that a factor in the factor model will lead to its indicators to cluster (all nodes connected to each other with strong edges) in the GGM. This result is in line with mathematical equivalences between factor models and network models of binary variables \cite{marsman2017introduction,kruis2,netpsych,marsman2015bayesian}. As there is only one unique inverse to $\pmb{\Sigma}$, there is only one unique GGM for every factor model. Conversely, however, one GGM may be equivalent to many different factor models (e.g., all possible rotations of $\pmb{\Lambda}$). 

Due to these equivalences, network modeling and factor modeling are closely connected. A natural first step in performing an exploratory factor analysis would be to estimate and draw a GGM model and investigate if the nodes cluster as would be expected by a factor model. Cluster-detection algorithms on the GGM could even be performed to investigate the number of factors to extract \cite{EGA}. Of note, however, is that many GGM estimation methods will always aim to estimate \emph{sparse} GGM (i.e., $\pmb{K}$ contains exact zeroes), which is not expected given a factor model (except when latent variables are orthogonal). As such, estimating a sparse network does not provide evidence that a latent variable model could not have generated the data \cite{boschlooCommentary,netpsych}. GGM modeling can further be used to augment factor analysis by modeling the latent variable variance--covariance matrix $\pmb{\Psi}$ or the residual variance--covariance matrix $\pmb{\Phi}$ as a GGM \cite{epskampPsychometrika}. Modeling $\pmb{\Psi}$ as a GGM leads to a \emph{latent network model}, which can be used in exploratory estimation of relationships between latent variables. Modeling $\pmb{\Phi}$ as a GGM leads to a \emph{residual network model}, which may be used to estimate factor models while local independence is structurally violated \cite{epskampPsychometrika,pan2017alternative}.

\paragraph{A note on spurious edges} 

When interpreting edges in the GGM as indicative of potential causal effects it is important to note that edges in a GGM may also result from latent variables. Such edges are termed \emph{spurious}, and cannot be accounted for unless the latent variable is explicitly modeled (e.g., by using the residual network model described above). The same problem occurs in exploratory DAG estimation, in which case a latent variable may induce a directed edge in the causal network. Furthermore, such spurious associations may arise in any statistical model, to the extent that unmeasured latent variables are involved. Here, the downside that GGM loses information on direction of effect turns into an upside: when an edge is indicative of a causal effect, GGMs do not retrieve the direction of effect, however, when an edge is spurious due to the influence of a latent variable, the GGM also does not introduce a strong causal hypothesis on what would happen under intervention.

\section{Estimating GGMs From Different Sources of Data}

\subsection{Data with Independent Cases}

A GGM can be estimated in datasets where cases can be assumed to be independent. Three common examples of such data are cross-sectional data, in which every subject is only measured once on a set of response items, aggregated data, in which only one mean score per variable per subject is included in the dataset, or $n = 1$ time-series data that feature large intervals between measurement occasions. In time-series data featuring shorter intervals, a GGM can be estimated as well; in this case, the network could be termed a contemporaneous network. However, as we argue in the next section on temporally ordered data, better methods exist that take temporal information into account in addition to modeling the contemporaneous effects in a GGM.

\subsubsection{Estimation}

In cross-sectional data analysis, only one observation per subject is available; thus, we cannot expect to estimate subject-specific means or GGM networks. It is typically assumed that the subjects all share the same distribution. That is,
\begin{equation*}
\pmb{y}_{P} \sim N\left(\pmb{0},  \pmb{\Sigma} \right),
\end{equation*}
in which $\pmb{y}_{P}$ denotes the random response of subject $P$ on all items. Similarly, in $n = 1$ time-series data we can make a similar assumption:
\begin{equation*}
\pmb{y}_{T} \sim N\left(\pmb{0},  \pmb{\Sigma} \right), 
\end{equation*}
in which $\pmb{y}_{T}$ denotes the random response of a subject on all items at time point $T$. In both cases, the full likelihood can be readily obtained, and the variance--covariance matrix $\pmb{\Sigma}$ can reliably be estimated using maximum likelihood estimation (MLE), least-squares estimation, or Bayesian estimation. 

\paragraph{Regularization} The MLE solution of $\pmb{K}$---the precision matrix encoding a GGM---can be obtained by standardizing the inverse sample variance--covariance as per Equation~\eqref{eq:parcor}. To obtain a sparse network, model search can be performed by iteratively adding and removing edges and fitting the corresponding GGM structure \cite{epskampPsychometrika}. In recent literature, it has become increasingly popular to use regularization techniques, such as penalized MLE, to jointly estimate model structure and parameter values \cite{costantini2015state, van2014new}. The \emph{least absolute shrinkage and selection operator} (LASSO; \citeNP{tibshirani1996regression}) has been shown to perform well in quickly estimating model structure and parameter estimates of a sparse GGM \cite{friedman2008sparse,meinshausen2006high,yuan2007model}. A particularly popular variant of LASSO is the \emph{graphical LASSO} (GLASSO; \citeNP{friedman2008sparse}), which directly penalizes elements of the inverse variance--covariance matrix  \cite{witten2011new,yuan2007model}. The GLASSO algorithm is useful as it is typically faster than other GGM estimation algorithms (which conduct multiple separate regressions and then combine the results using Equation~\ref{dynamics:eq:regToGGM}), and requires only an estimate of the variance--covariance matrix rather than raw data \cite{primerpaper}. LASSO utilizes a tuning parameter which can be chosen in a way that optimizes cross-validated prediction accuracy or that minimizes information criteria such as the extended Bayesian information criterion (EBIC; \citeNP{chen2008EBIC}). Estimating a GGM with the GLASSO algorithm in combination with EBIC model selection has been shown to work well in retrieving the true network structure \cite{qgraphsims,foygel2010extended}. For an introduction to this methodology aimed at empirical researchers, we refer the reader to \citeA{primerpaper}. 

\paragraph{Software} Several software packages allow for GGM estimation as described above. MLE can be performed in any programming language and in many statistical programs by inverting and subsequently standardizing the sample variance--covariance matrix. In the open-source statistical programming language R \cite{R}, automated procedures have been implemented in the \emph{corpcor} package \cite{corpcor} and the \emph{qgraph} \cite{jssv048i04} package. The qgraph package also supports thresholding via significance testing or false discovery rates.  The GLASSO algorithm is implemented in the \emph{glasso} \cite{glasso} and \emph{huge} \cite{huge} packages. EBIC-based tuning parameter selection using the \emph{glasso} package has been implemented in the \emph{qgraph} package. The \emph{huge} package also allows for selection of the tuning parameter using cross validation or EBIC. The \emph{parcor} package \cite{parcor} implements other LASSO variants to estimate the GGM. The \emph{BDgraph} package \cite{mohammadi2015bdgraph} implements a Bayesian method to estimate the undirected structure. Finally, fitting an estimated GGM to data can be done in the R packages \emph{ggm} \cite{ggm} and \emph{lvnet} \cite{epskampPsychometrika}.

\subsection{Temporally Ordered Data of a Single Subject}

In line with a call for more intraindividual and person-based research \cite{molenaar2004manifesto}, an increasingly popular form of data pertains to $n=1$ time series, in which a single individual is measured repeatedly over a period of time. One such situation is in clinical practice \cite{KroezeR,contempraneous}, where a patient can be measured several times per day over a period of a few weeks. We will limit our discussion to data obtained in a relatively short time-frame so that we can reasonably assume the model will remain stable over time. Then, we can apply the methodology above to obtain a GGM for the $n=1$ data. However, such an analysis does not take temporal ordering of data into account (i.e., relationships between measurement occasions) and only investigates contemporaneous relationships between variables (e.g., within the same measurement occasion). This is important for several reasons. First, valuable information, especially in the context of dynamical relationships, might be contained at the temporal level rather than at the contemporaneous level. Second, not taking temporal ordering into account might bias the estimated contemporaneous relationships (see Section 4 of the supplementary materials). For example, if one variable causes itself and another variable at the next time, then not taking temporal ordering into account turns that variable into a latent cause, which would produce an edge in the GGM. Third, temporal information is needed when constructing the joint likelihood over time (e.g., to obtain the information retained in a system over time; \citeNP{roadahead, quax2013information}). Finally, temporal information can aide in distinguishing reciprocal and cyclic effects by regarding these as acyclic effects unfolding over time.

\paragraph{Vector Auto-regression}  The simplest way to deal with temporal ordering of cases is to incorporate the effect between consecutive measurements \cite{shumway2010time,hamilton1994time,chatfield2016analysis}. This is called a Lag-$1$ model because it includes both measurements at the current time point $t$ as well as measurements from the previous time point $t-1$. We will focus our discussion on Lag-$1$ models, noting that everything below also generalizes to more complicated models (e.g., Lag-2 models). In intraindividual analysis, VAR (\citeNP{brandt2007multiple, rosmalen2012revealing}) has gained substantive footing in visualizing temporal information through networks. The lag-1 VAR model can be denoted as a regression model on the previous measurement occasion:
\begin{align}
\pmb{y}_{t} &= \pmb{B} \pmb{y}_{t-1}  + \pmb{\varepsilon}_{t} \nonumber\\
\pmb{\varepsilon}_{T} &\sim N( \pmb{0}, \pmb{\Theta}). \label{typicalVAR}
\end{align}
The model matrix $\pmb{B}$ encodes temporal predictive effects from variables on variables in the next measurement occasion, and can be used to obtain a directed network, which we term the \emph{temporal network}. The variance--covariance matrix $\pmb{\Theta}$ can be inverted ($\pmb{K}^{(\pmb{\Theta})} = \pmb{\Theta}^{-1}$) to obtain a GGM modeling effects within the same measurement occasion, after controlling for temporal effects. These can be displayed again as a network, which we term the \emph{contemporaneous network}.

\paragraph{Temporal networks} Temporal networks, encoded by $\pmb{B}$, have grown popular in recent psychological literature (e.g., \citeNP{bringmann2013network,bringmann2015revealing,wigman2015exploring,fionneke,snippe2017impact,klippel2017cascade}). A temporal network is formed by combining a lagged variable $y_{t-1}$ and current variable $y_{t}$ into a single node, connected with directed edges which are weighted according to the regression parameters contained in $\pmb{B}$.\footnote{Note, in graph theory it is common to encode a network using a \emph{weights matrix} in which the row indicates the node of origin and the column indicates the row of destination. As such, to obtain the directed weights matrix to draw a temporal network $\pmb{B}$ needs to be transposed.} Thus, an edge in the temporal network indicates that a node predicts another node (or itself in the common case of self-loops) at the next measurement occasion, after controlling for all other variables at the previous measurement occasion. Temporal prediction is central to the concept of  \emph{Granger causality} in the economic literature \cite{eichler2007granger,granger1969investigating}, and it satisfies at least the temporal requirement for causation (i.e., the cause must precede the effect). Temporal networks may thus highlight potential causal pathways. While temporal networks are typically cyclic, they can also be interpreted as summarizing a DAG unfolding over time.

\paragraph{Contemporaneous networks} In addition to temporal effects, VAR analyses also include contemporaneous effects, which can be modeled as a GGM. We will term this modeling framework (a VAR model with contemporaneous effects explicitly modeled and portrayed as a GGM) \emph{graphical VAR} (GVAR; \citeNP{wild2010graphical}).\footnote{\citeA{wild2010graphical} do not use the term graphical VAR in the exact same way we do, and use it more to refer to graphical modeling in a VAR framework, including structural VAR. We use the term here as described because having an explicit term helps in contrasting GVAR from, e.g., structural VAR.} A useful equivalent way to denote a GVAR model is by using a conditional Gaussian distribution:
\begin{align*}
\pmb{y}_{T} \mid \pmb{y}_{T-1} = \pmb{y}_{t-1} &\sim N\left( \pmb{B} \pmb{y}_{t-1} , \pmb{\Theta}\right).
\end{align*}
Which is equivalent to Equation~\eqref{typicalVAR}. Now, it becomes evident that if consecutive cases can be assumed to be independent, and thus $\pmb{B} = \pmb{O}$, the GVAR model is exactly the same as the GGM model described above for independent cases. Thus, the GVAR model can be seen as a generalization of the GGM model to temporally ordered data. GVAR only differs from regular VAR in that the contemporaneous structure is modeled and represented as a GGM, instead of being saturated. This leads to a strikingly different interpretation of the VAR model; the VAR model can be seen as an inclusion of temporal effects on a GGM.

\begin{figure}
\centering
\includegraphics[width=1\linewidth]{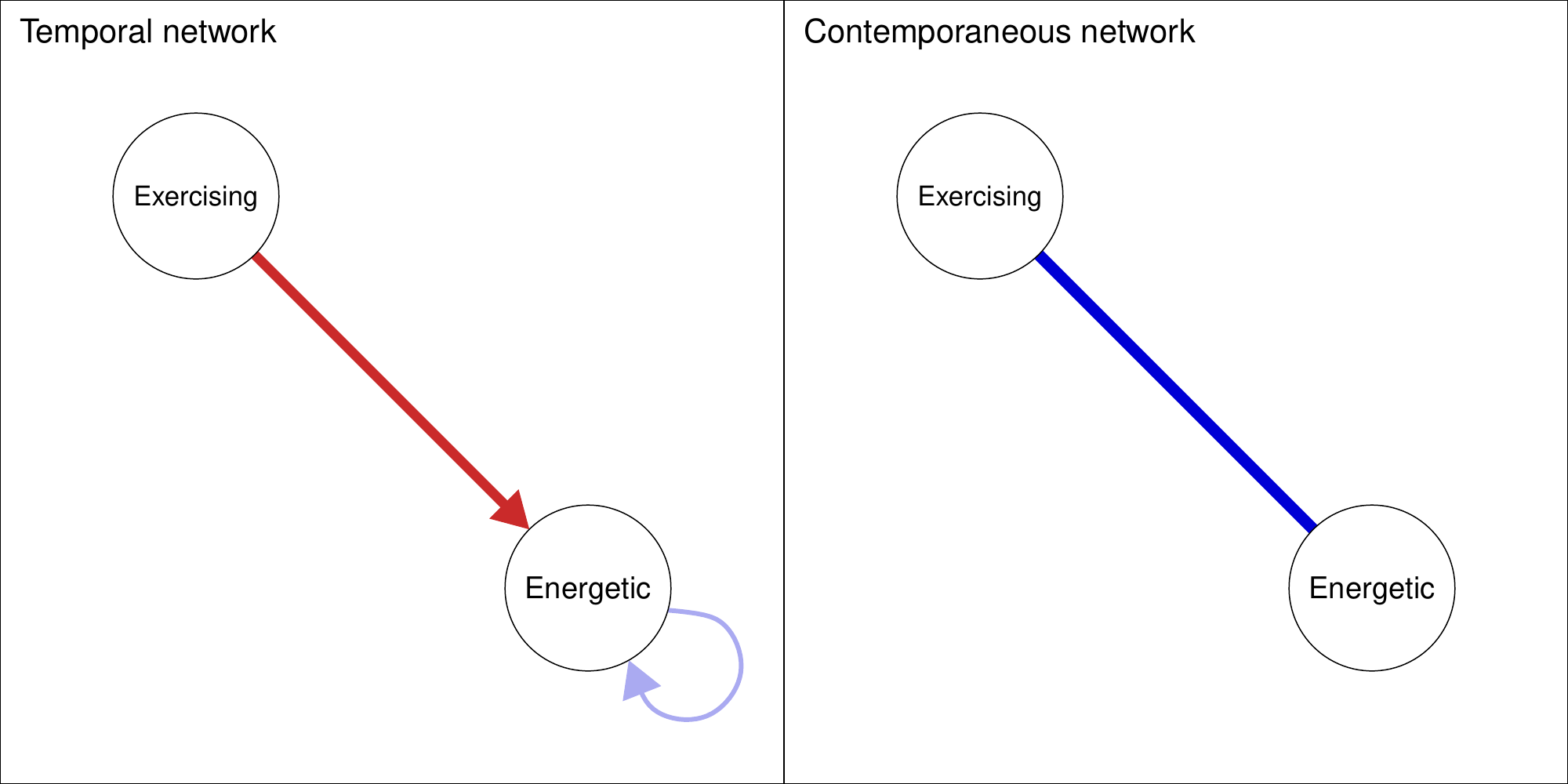}
\caption{A hypothetical example of two network structures obtained from a GVAR analysis. The network on the left indicates the temporal network, demonstrating that a variable predicts another variable at the next time point. The network on the right indicates the contemporaneous network, demonstrating that two variables predict each other at the same time point.}
\label{dynamics:fig:VAR}
\end{figure}

\paragraph{Temporal and contemporaneous information} Figure~\ref{dynamics:fig:VAR} shows a hypothetical example of the two network structures obtained in a GVAR analysis and shows how they might plausibly differ. The left panel shows the temporal network. The self-loop shows that whenever the subject in question felt energetic (or tired) this person also felt more (or less) energetic in the next measurement. The temporal network also shows us that after exercising, this person felt less energetic. The contemporaneous network in the right panel shows a plausible reverse relationship: Whenever this person exercised, he or she felt more energetic in the same measurement occasion. In psychology, there will likely be many causal relationships that occur much faster than the lag interval of a typical ESM study; in this case, these pathways will be captured in the contemporaneous network. For example, if someone is experiencing bodily discomfort, that will immediately negatively affect that person's ability to enjoy him or herself \cite{contempraneous}. Especially when the measurement is on blocks of time (e.g., ``since the last measurement did you feel ...''), such effects are likely to be caught in the contemporaneous network.

\subsubsection{Estimation}

Estimating saturated (fully connected temporal and contemporaneous networks) GVAR models is straightforward. First, one needs to estimate temporal effects of a regular VAR model by performing multivariate multiple regression of all variables on the previous measurement occasion,
\begin{align*}
\pmb{y}_{t} &= \pmb{B} \pmb{y}_{t-1} + \pmb{\varepsilon}_{t},
\end{align*}
or by estimating univariate models for every variable,
\begin{align*}
y_{ti} &= \pmb{\beta}_i \pmb{y}_{t-1} + \varepsilon_{ti}, 
\end{align*}
in which $\pmb{\beta}_i$ denotes the $i$th row of $\pmb{B}$. Next one can invert the variance--covariance matrix of the residuals to obtain a GVAR model. Step-wise model selection in latent network models \cite{epskampPsychometrika} can also be used to estimate sparse GVAR models. Missing data can be handled in default ways of SEM or regression models (e.g., listwise deletion and full-information maximum likelihood), or by using more sophisticated techniques such as Bayesian estimation \cite{schuurman2016comparison} or the Kalman filter \cite{harvey1990forecasting,kim1999state}. 

\paragraph{Novel estimation methods} A promising recent method for estimating VAR models is the Bayesian dynamical SEM implementation in version 8 of Mplus \cite{muthen2007mplus,asparouhov2016dynamic}, which includes handling of missing data, measurement invariance and latent variables. Mplus can be used to estimate saturated GVAR models, and to perform model selection in the temporal network of a GVAR model. Model selection in the contemporaneous network of a GVAR model is not yet implemented, but credibility intervals around contemporaneous effects can be obtained by manually inverting each sampled residual variance--covariance matrix (these can be stored using the BPARAMETERS option).

When estimating GVAR models, regularization methods can be used similar to the estimation of GGMs on non-temporally ordered data. \citeA{abegaz2013sparse} proposed to apply LASSO estimation to jointly estimate the temporal and contemporaneous network structures using the multivariate regression with the covariance estimation (MRCE) algorithm described by \citeA{rothman2010sparse}. MRCE involves iteratively optimizing $\pmb{B}$, using cyclical-coordinate descent, and $\pmb{K}^{(\pmb{\Theta})}$, using the GLASSO algorithm \cite{friedman2008sparse,GLASSO}. EBIC model selection can be used to obtain the best performing model. This methodology has been implemented in two open source R packages: \emph{sparseTSCGM} \cite{SparseTSCGM},  which aims to estimate the model on repeated multivariate genetic data, and \emph{graphicalVAR} \cite{graphicalVAR}, which was designed to estimate the model on the psychological data of a single subject. The \emph{graphicalVAR} package also allows for unregularized multivariate estimation.

An alternative to estimating GVAR models is to estimate \emph{structural VAR} (SVAR; \citeNP{chen2011vector}) models, also called \emph{unified SEM} (\citeNP{gates2010automatic}). In SVAR, the contemporaneous effects are modeled using a directed network instead of an undirected network. The sparsity of the undirected GVAR contemporaneous network corresponds in the same way to the sparsity of the directed contemporaneous network in SVAR as how the GGM corresponds to causal models (edges arise in the GGM due to edges in the causal network or conditioning on common effects). The temporal SVAR network is sparser than the temporal GVAR network, as contemporaneous mediators can be controlled for in SVAR but not in GVAR. A saturated SVAR model can be obtained by using regressions on the previous time-point as mentioned above, followed by transforming the contemporaneous variance--covariance matrix (e.g., by using a Cholesky decomposition on its inverse; \citeNP{lutkepohl2005}) and subsequently transforming the temporal effects to take contemporaneous mediators into account. This technique of obtaining an SVAR model leads to multiple solutions \cite{beltz2016dealing}. Step-wise model selection can also be used to estimate sparse SVAR, for example by using model selection in (unified) SEM \cite{gates2010automatic} or Bayesian dynamical SEM models \cite{muthen2007mplus,asparouhov2016dynamic}.

\subsection{Temporally Ordered Data of Multiple Subjects}

A type of data that is increasingly common due to the emergence of ESM studies is time series of multiple subjects (e.g., \citeNP{ bringmann2013network, bringmann2015revealing, mottus2016within, schmiedek2010hundred, wigman2015exploring}). Such datasets pose a promising gateway to study both intraindividual dynamics and between-subjects overlap as well as their differences. Here, we assume that the number of time points might differ per person and that measurement occasions are nested in people. We can model the temporal data of every person with an individual GVAR model:
\begin{align*}
\pmb{y}_{[t,p]} &= \pmb{\mu}_p + \pmb{B}_{p} \left(\pmb{y}_{[t-1,p]} - \pmb{\mu}_{p} \right) + \pmb{\varepsilon}_{[t,p]} \\
\pmb{\varepsilon}_{[T,p]} &\sim N( \pmb{0}, \pmb{\Theta}_{p}) \\
\pmb{\Theta}_{p}^{-1} &= \pmb{K}^{(\pmb{\Theta})}_{p},
\end{align*}
in which $\pmb{\mu}_{p}$ indicates the stationary mean vector of subject $p$ (which enters the model because we can no longer assume within-subject means are zero without loss of generality), $\pmb{B}_{p}$ encodes the person-specific temporal network, and $\pmb{K}^{(\pmb{\Theta})}_{p}$ encodes the person-specific contemporaneous GGM. 

\paragraph{Multilevel modeling} To gain insight in the general network structure over subjects we can investigate the individual networks at a second level. Doing so is termed \emph{multi\-level modeling}, explained in more detail in section 2.1 of the supplementary materials. Let $\pmb{B}_{*}$ and $\pmb{K}^{(\pmb{\Theta})}_{*}$ encode the expected temporal and contemporaneous network when selecting a person at random. Furthermore, we can assume without loss of generality that data are grand-mean centered. We then obtain:
\begin{align*}
\mathcal{E}\left(\pmb{\mu}_P\right) &= \pmb{0} \\
\mathcal{E}\left(\pmb{B}_P\right) &= \pmb{B}_{*} \\
\mathcal{E}\left(\pmb{K}^{(\pmb{\Theta})}_P\right) &= \pmb{K}^{(\pmb{\Theta})}_{*}.
\end{align*}
Here, $\pmb{B}_{*}$ and $\pmb{K}^{(\pmb{\Theta})}_{*}$ now encode the average parameters in the population: the \emph{fixed effects}. Deviations from these fixed effects, such as $\pmb{B}_p - \pmb{B}_{*}$, are often called \emph{random effects}. Besides the individual network structures, researchers often aim to estimate the structure and parameters of these fixed effects because these tell us something about the average intraindividual effect. Researchers also aim to estimate the variance--covariance structure of the random effects because it tells us something about individual differences \cite{bringmann2013network}.

The random effects can be modeled by assuming a second level normal distribution on all the parameters. This can be complicated, however, especially when modeling partial correlation coefficients in such a way (e.g., any hierarchical model for $\pmb{K}^{(\pmb{\Theta})}$ needs to take into account that this matrix must remain positive definite). The interpretation of, for example, correlations between different temporal or contemporaneous edges is also difficult. Therefore, we only focus here on a subset of the parameters where we can easily interpret the second-level model: the mean structure. As a result, if a multivariate normal is assumed for all parameters, then it is also assumed for the marginal distribution of the means---regardless of other parameters:
\begin{equation*}
\pmb{\mu}_P \sim N\left(\pmb{0}, \pmb{\Omega}\right).
\end{equation*}
Again, we can invert the variance--covariance matrix to obtain a GGM,
\begin{equation*}
\pmb{K}^{(\pmb{\Omega})} = \pmb{\Omega}^{-1},
\end{equation*}
which we will term the \emph{between-subjects network}, a network between stationary means of different subjects.\footnote{Of note, it is also possible to invert and standardize the full random effects variance--covariance matrix, which would lead to different between-subjects relationships between the means as well (partial correlations after conditioning on other means and all other between-subject parameters such as edges). We do not do that here as (a) such a network is hard to interpret, and (b) most estimation methods we mention do not return the full random effects variance--covariance matrix (especially the correlations between temporal and contemporaneous edges are hard to obtain).} As such, estimating the GVAR model on $n >  1$ time-series analysis allows for the separation of variance into three distinct network structures: temporal networks, contemporaneous networks, and the between-subjects network.

\subsubsection{Estimation} In this section, we outline several different ways in which individual network structures as well as fixed effect network structures may be estimated. We first discuss applying the methodology of estimating $n = 1$ GVARs discussed above to both pooled data as well as data of each subject separately, followed by a discussion of different multilevel estimation procedures that take clustering of the data into account. An overview of these methods is also included in Table~\ref{VARtable}.

\paragraph{Pooled and individual LASSO estimation} First, we can estimate a GVAR model for every subject to obtain subject-specific estimates for the temporal and contemporaneous networks. Similarly, we can estimate fixed-effects networks by estimating a GVAR model on the entire within-subjects centered dataset, using the sample means of every subject on every variable as a plug-in for the within-subject means. Consequently, we can estimate the between-subjects network by estimating a GGM on the sample means of each subject on all variables. We can readily apply the LASSO regularization methods described earlier for this purpose: the methodology outlined by \citeA{abegaz2013sparse} to estimate temporal and contemporaneous networks and the methodology outlined by \citeA{foygel2010extended} to estimate the between-subjects GGM. We term this framework \emph{pooled and individual LASSO estimation} and have implemented it in the R package graphicalVAR \cite{graphicalVAR}. The performance of pooled and individual LASSO estimation is assessed in simulations reported in section 3 of the supplementary materials.

\paragraph{Multilevel estimation} The second and third procedures described in Table~\ref{VARtable} make use of multi\-level modeling \cite{hamaker2012researchers}. Two main benefits of this approach are (1) instead of estimating the VAR model in each subject, only the fixed effects and variance--covariance of the random effects need to be estimated, and (2) afterwards, estimates of subject-specific parameters can be obtained, which are somewhat pulled together (termed \emph{shrinkage}). Shrinkage allows the estimation of the model for one subject to borrow information from other subjects. Multi\-level estimation can be performed by specifying the multivariate model using hierarchical Bayesian Monte-Carlo sampling methods or by integrating over the distribution of the random effects \cite{gelman2006data,schuurman2016comparison}. 

\paragraph{Multivariate Bayesian multilevel} Bayesian multivariate estimation has proven to be powerful in estimating multivariate multilevel models, especially given its flexibility in adding measurement error, latent variables and in handling missing data \cite{schuurman2015incorporating}. Recently, the dynamic SEM methodology implemented in Mplus version 8 \cite{muthen2007mplus,asparouhov2016dynamic} has made estimation of multivariate multi\-level VAR models much faster and more user-friendly than other Bayesian software routines. Specifying a temporal VAR model with correlated random effects is straightforward and relatively fast to compute with a moderate number of variables (e.g., 6). At the time of writing, Mplus does not return partial correlations by default, but these can be obtained by using the BPARAMETERS option and manually inverting the sampled variance--covariance matrices. Mplus allows for specifying random effects on the contemporaneous covariances and thus, by extension, allows for estimating random contemporaneous networks in addition to random temporal networks. Specifying such a model can be done by specifying dummy latent variables for the residual covariance between each pair of variables (a prior guess on the sign of the covariance is needed). Doing so, however, can significantly increase computation length especially when all random effects are allowed to correlate. To facilitate estimation, we have implemented a function generating Mplus code for a multi\-level GVAR model and subsequently running the model using the \emph{MplusAutomation} package \cite{MplusAutomation} in version 0.4 of the \emph{mlVAR} package, which can be called using \verb|estimator = "Mplus"| and requires the Mplus program to be installed.

\paragraph{Two-step multilevel VAR} A downside of multivariate estimation is that the number of random effect covariances to be estimated increases quadratically with the number of variables. Forcing random effects to be uncorrelated helps, but places strict assumptions on the model. \citeA{bringmann2013network} proposed to estimate multi\-level VAR models using univariate models instead, using a frequentist estimation procedure. In this work, the multi\-level VAR model is estimated by sequentially estimating univariate multi\-level regression models of one variable given all lagged variables. Doing so ignores several correlations of random effects because many parameters are not estimated in the same model, simplifying the analysis: only correlations between incoming edges to the same node and the intercept of that node are included in an univariate model. This method scales up well to approximately eight variables when estimating correlated random effects and around $20$ variables when estimating orthogonal random effects (or by using a moving window approach; \citeNP{bringmann2015revealing}). Of note, when specifying orthogonal random effects not all random effects are assumed to be uncorrelated, merely the ones used in the same univariate model.

\begin{table*}
	\caption{Three Methods of Estimating GVAR Models With $n >  1$ Subjects}
	\centering
	{\footnotesize
		\begin{tabular}{l p{4cm} p{4cm} p{4cm}}
			\toprule
			& Pooled and individual LASSO estimation & Bayesian multi\-level & Two-step frequentist multi\-level  \\
			\midrule
			Software & \emph{graphicalVAR} \cite{graphicalVAR}; \emph{sparseTSCGM} \cite{SparseTSCGM}.  & \emph{MPlus} 8 \cite{muthen2007mplus,asparouhov2016dynamic}; \emph{mlVAR} (wrapper around Mplus). & \emph{mlVAR} \cite{mlvar}. \\[0.5cm]
			Estimation  &  (1) Joint multivariate LASSO estimation with EBIC model selection \cite{abegaz2013sparse} of within-subjects centered data to obtain fixed effects temporal and contemporaneous networks. (2) GLASSO algorithm with EBIC model selection \cite{foygel2010extended} on sample means of subjects to obtain between-subject network. (3) Step (1) repeated for each individual dataset to obtain subject-specific networks. &  MCMC sampling from multivariate hierarchical model (e.g., \citeNP{schuurman2016comparison}). &  (1) Sequential univariate multi\-level regression models on previous measurement (similar to \citeNP{bringmann2013network}), with within-subject centered lagged variables as within-subjects level predictors and sample-means of all other variables as between-subjects predictor. (2) Sequential multi\-level regression models using the residuals of (1): residuals of one variable are predicted by residuals of all other variables in the same measurement occasion.  \\[0.5cm]
			Pros & Fast estimation of fixed effects; scales up well to large numbers of nodes; model selection in individual networks; temporal and contemporaneous networks obtained in the same analysis. & Borrowing information in individual network estimation from other subjects; all model parameters and random-effect (co)variances can be estimated; credibility intervals can be obtained for edges and descriptive statistics (e.g., centrality; density); advanced extensions such as measurement error and latent variable modeling possible; powerful handling of missing values. &  Borrowing information in individual network estimation from other subjects; scales up well to ~8 nodes (correlated random effects) or ~20 nodes (orthogonal random effects); many random effect variances correlations can be estimated; fast estimation of individual networks.  \\[0.5cm]
			Cons &  Fixed effects estimated on pooled data; Subject specific networks estimated without borrowing information from other subjects (no multi\-level structure); between-subjects network estimated in a different model; very slow to estimate subject-specific networks; poor handling of missing values. & Relatively slow estimation, especially in higher dimensional models; no model selection (thresholding possible via credibility intervals); complicated to estimate contemporaneous random effects. &  Slow estimation in larger datasets; no model selection (fixed effects can be thresholded using significance); combination of many different models; does not scale up well past 20 nodes; poor handling of missing values.  \\
			\bottomrule
		\end{tabular}
	}
	\small\textit{Note}. {The software listed only concerns user-friendly automated software because all these models could readily be implemented in most programming languages or Bayesian sampler packages.}
	\label{VARtable}
\end{table*}

The methodology of \citeA{bringmann2013network} does not estimate contemporaneous or between-subjects networks. To this end, we extended the algorithm in a framework we term \emph{two-step multi\-level VAR}. The details of this estimation procedure are explained in Section 2 of the supplementary materials. In short, we extend the methodology of \citeA{bringmann2013network} by within-subject centering and by adding subject sample means as between-subjects predictors (as discussed by e.g., \citeNP{hoffman2009persons,curran2011disaggregation,hamaker2014center}). This allows us to estimate between-subjects networks by collecting regression coefficients as in Equation~\eqref{dynamics:eq:regToGGM} and symmetrizing the resulting matrix.\footnote{Standardizing regression parameters from nodewise multi\-level models to partial correlation coefficients does not lead to perfectly identical estimates.} 
In a second step, we take the residuals of the first analysis and again perform sequential univariate multi\-level regression models to predict each residual from all other residuals in the same measurement occasion.  Again, these can be collected, as in Equation~\eqref{dynamics:eq:regToGGM}, and symmetrized to obtain contemporaneous networks. 
Networks can be thresholded by removing all effects that are not significant. For the between-subjects and contemporaneous networks, this results in two $p$-values for every edge---either both can be required to be significant (``and''-rule) or an edge can be included if one of the two $p$-values is significant (``or''-rule). Using the ``and''-rule means erring more on the side of caution (sparser network), whereas using the ``or''-rule means erring more on the side of discovery. We have implemented two-step multi\-level VAR in the \emph{mlVAR} package, which can be called using \verb|estimator = "lmer"| (the default).

\paragraph{Choosing the estimation method} The choice of which estimator to use is not trivial and depends on the interests of the researcher. In Table~\ref{VARtable} we list some pros and cons of each of the methodologies. In particular, multi\-level estimation can be very complicated and is harder in high-dimensional settings. Assuming  normally distributed parameters can also be problematic because doing so imposes that subjects cannot differ on the structure of the networks, merely on their parameterization. When a parameter (e.g., a temporal edge) is zero in some subjects but nonzero in others, then this parameter cannot be normally distributed (the distribution would peak at $0$). Therefore, it is currently hard to estimate differently structured individual networks (different edges set to be exactly $0$ between subjects) in multi\-level estimation. Nonetheless, multi\-level estimation particularly shines in that when estimating an individual network, researchers can borrow information from other subjects. We have performed simulation studies to assess the performance of the two proposed methods in this paper: pooled and individual LASSO estimation and two-step multi\-level VAR. We report the results of these studies in Section 3 of the supplementary materials, which shows that both methods adequately detect the true fixed-effect network structures with increasing sample size. Having more time points per subject helps to estimate the contemporaneous and temporal networks, and having more subjects helps to estimate the between-subject networks. Two step multi\-level VAR performs well in estimating intraindividual networks when the number of observations is low, but does not perform subject-specific model selection: all estimated intra-individual networks are saturated and contain all edges. Pooled and aggregated LASSO estimation does estimate the structure of intraindividual networks, but performs poorer in intraindividual parameter estimation with fewer observations as no information is borrowed from other subjects.

\paragraph{GIMME}  Finally, when analyzing $n > 1$ data, another option is to estimate SVAR models instead. A promising estimation procedure to estimate such models over many individuals, while dealing with potential heterogeneity, is ``group iterative multiple model estimation'' (GIMME; \citeNP{gates2012group}), which is implemented in R using the \emph{gimme} package \cite{gimme}. In GIMME, no multi\-level structure is imposed and subject-specific networks are allowed to differ in structure. Information from other subjects is borrowed, however, in that the structure of individual networks can be based on other subjects (e.g., an edge can be included because it is present in many other subjects). No shrinkage is induced on the parameter estimates that are nonzero (as would be the case in multi\-level or hierarchical Bayesian modeling). A variant of GIMME that estimates the GVAR or a combination of structural and GVAR models has not yet been developed, and we note that this may be a promising avenue for further research.

\section{Interpreting GGMs Estimated from Between-subjects Data}

This paper describes the estimation of network models on data from different subjects (both cross-sectional as well as person-wise average scores). Cross-sectional network modeling is often criticized for inappropriately taking cross-sectional results to be reflective of within-person causal processes (e.g., \citeNP{bos2016group,fionneke}), as it can be shown that such results will not equal within-person processes except under strong assumptions \cite{molenaar2004manifesto}. To this end, this section discusses the interpretation of GGMs estimated from data of different subjects. We first argue that cross-sectional data may be interpreted as a between-subjects analysis, assuming that between-subjects variance is dominant, and next discuss the interpretation of potential causal effects at the between-subjects level.

\subsection{Cross-Sectional Data Analysis}

\paragraph{Within- and between-subjects variation}  A type of data to which the GGM is currently often applied is data belonging to multiple subjects that are all measured only once (e.g., \citeNP{isvoranu, van2015association}). Such a dataset is often termed cross-sectional data, and such an analysis is often termed a between-subjects analysis. However, the term between-subjects analysis might not  be warranted, as it is difficult to distinguish between within-subject variation around an individual's stable mean and between-subject variation of such stable within-subject means using only cross-sectional data \cite{hamaker2012researchers}. It is well known that subjects might respond differently when measured multiple times \cite{lord1968statistical}. As such, the single observation per subject leads to the time point and the subject being random: $\pmb{y}_{[T,P]}$. We might make the argument that two distinct sources of variation cause the outcome \cite{bolger2013intensive}. Repeated measures of a subject (here $p$) are distributed according to an unique within-subject model:
\[
  \pmb{y}_{[T,p]} \sim N(\pmb{\mu}_{p}, \pmb{\Sigma}_p),
\]
That is, of a particular response, the subject's score is a composite of the average stationary score $\pmb{\mu}_{p}$ and random deviation.\footnote{Section 4 of the supplementary materials show that when consecutive cases ($t$ and $t+1$) are assumed dependent, such a zero-order network may result from a mixture of temporal and contemporaneous effects as described above. The discussion here does not concern estimation of model parameters and hence does not require an assumption of independence of cases.} These average stationary scores also differ in the population. Thus, we need to model the average stationary scores of a random subject $P$ with a separate distribution:
\[
  \pmb{\mu}_{P} \sim N(\pmb{0}, \pmb{\Omega}),
\]
in which we can assume, without loss of generality, an overall mean of $\pmb{0}$. We can invert the variance--covariance matrix $\pmb{\Omega}$ to obtain a GGM:
\begin{equation*}
\pmb{K}^{(\pmb{\Omega})} = \pmb{\Omega}^{-1}.
\end{equation*}
This GGM corresponds to a between-subjects network. The matrix $\pmb{\Sigma}_p$ can also be inverted and standardized to a GGM to obtain a within-subject network:
\[
\pmb{K}_p = \pmb{\Sigma}_p^{-1}.
\]
We will term this network a \emph{within-subjects network}.

\paragraph{The value of a cross-sectional analysis} It is immediately clear that with only one response per subject we cannot hope to estimate subject-specific variance-covariance matrices $\pmb{\Sigma}_p$ (and as a result individual GGMs). Moreover, even if we assume that within-subject effects are equal across subjects (denoted with $\pmb{\Sigma}_*$ below), this still leaves us without an estimable model because $\pmb{\mu}$ is also assumed to be normally distributed. The co-variation between responses thus becomes an unidentified blend of $\pmb{\Sigma}_*$ and $\pmb{\Omega}$: $A$ and $B$ may correlate in cross-sectional data because people who score on average high on $A$ also score on average high on $B$ (trait-level variation in $\pmb{\Omega}$), or because when people deviate from their average on $A$ they also tend to deviate from their average on $B$ (state-level variation in $\pmb{\Sigma}_*$). Even when within- and between-subjects effects are assumed not to correlate, the GGM estimated on such data becomes
\begin{equation*}
\pmb{K} = \left(\pmb{\Sigma}_* +  \pmb{\Omega}  \right)^{-1},
\end{equation*}
which is not a simple function of the between-subjects GGM and the within-subjects GGM. Only when no short-term within-subject variation, $\pmb{\Sigma}_* = \pmb{O}$, or no between-subjects variation, $\pmb{\Omega} = \pmb{O}$, is assumed does the cross-sectional GGM correspond exactly to one of the two networks.

Cross-sectional data analysis thus cannot disentangle between-subjects relationships from short term within-subjects relationships \cite{hamaker2012researchers}. 
For example, cross-sectional analysis cannot distinguish whether or not fatigue and concentration correlate because whenever people feel fatigued they also concentrate poorly (a within-subjects effect) or because people who are on average fatigued also tend to concentrate poorly on average (a between-subjects effect). However, preliminary simulation results show that the resulting cross-sectional GGM generally does not contain edges that are not present in either the within- or between-subjects network \cite{crosssection}. Depending on the ratio of within-to-between person variance, the cross-sectional analysis will pick up the within-subject network, the between-subject network, or a mixture of the two. As such, if one assumes between-subject variance to be dominant, the cross-sectional results may be interpreted as mainly reflecting between-subjects relations.

\paragraph{Cross-sectional analysis as between-subjects analysis} An important consideration is that a typical cross-sectional questionnaire or interview is vastly different than a typical ESM questionnaire, and many cross-sectional studies aim to measure variables that are more stable over time and for which a time-series analysis might not make sense. Good examples of this are recent network analyses in the area of schizophrenia \cite{isvoranua, isvoranu}, in which the impact of environmental factors (e.g., childhood trauma, urbanization) on psychotic symptoms and general psychopathology was studied. Such variables do not vary much over time; therefore, a cross-sectional analysis seems more suitable here. Other examples are questionnaires asking participants to rate symptoms over a period of several weeks or to describe themselves as ``I am a person who\ldots'' In such cases, the cross-sectional network may be interpreted as a between-subjects network, of which we discuss the interpretation below. Note that some research indicates that state-level variance (how a person feels at the moment) \emph{does} influence self-reported scores given trait-level (how a person feels on average) instructions \cite{brose2013affective}, indicating that this interpretation of cross-sectional results should be taken on a case-by-case basis depending on the topic studied. 

\subsection{Within- and between-subjects effects}

In contrast to prior work on multi\-level VAR modeling (e.g., \citeNP{bringmann2013network,bringmann2015revealing,wigman2015exploring,pe2015emotion}), in this paper between-subjects effects are conceptualized in addition to the within-subjects effects in a separate GGM. Furthermore, in contrast to prior work,  cross-sectional networks are not interpreted to be reflective of within-subject effects, but rather to potentially reflect a between-subjects structure, assuming that observed scores are not dominated by state-like variance (but see \citeNP{brose2013affective}). This raises the question on how such models could be interpreted. In particular, if edges in the GGM are interpreted as generating hypotheses to potential causal pathways, the question is raised how such causal effects can occur at the between-subjects level. This section therefore discusses the topic of causation at the between-subjects level. Here, we interpret the stationary means as being locally stationary: the average of a subject in a relatively short time span of measurement (e.g., a few weeks). As such, we do not interpret the mean vector $\pmb{\mu}_P$ as a lifetime average. Instead, we assume it could change, potentially due to experimental intervention. As a result, we argue that the between-subjects network can also be indicative of potential causal pathways---regardless of whether it is estimated from a cross-sectional interview concerning variables that are not expected to vary much over time or obtained from estimating the means from time-series data. To simplify the argumentation below, we do not discuss separate temporal and contemporaneous networks but only general within-subjects networks (a GGM of within-subject data without taking temporal ordering into account).

\begin{figure}
\captionsetup[subfigure]{justification=centering}
\centering
\begin{subfigure}{1\linewidth}
\centering
	  	\includegraphics[width=1\linewidth,page=1]{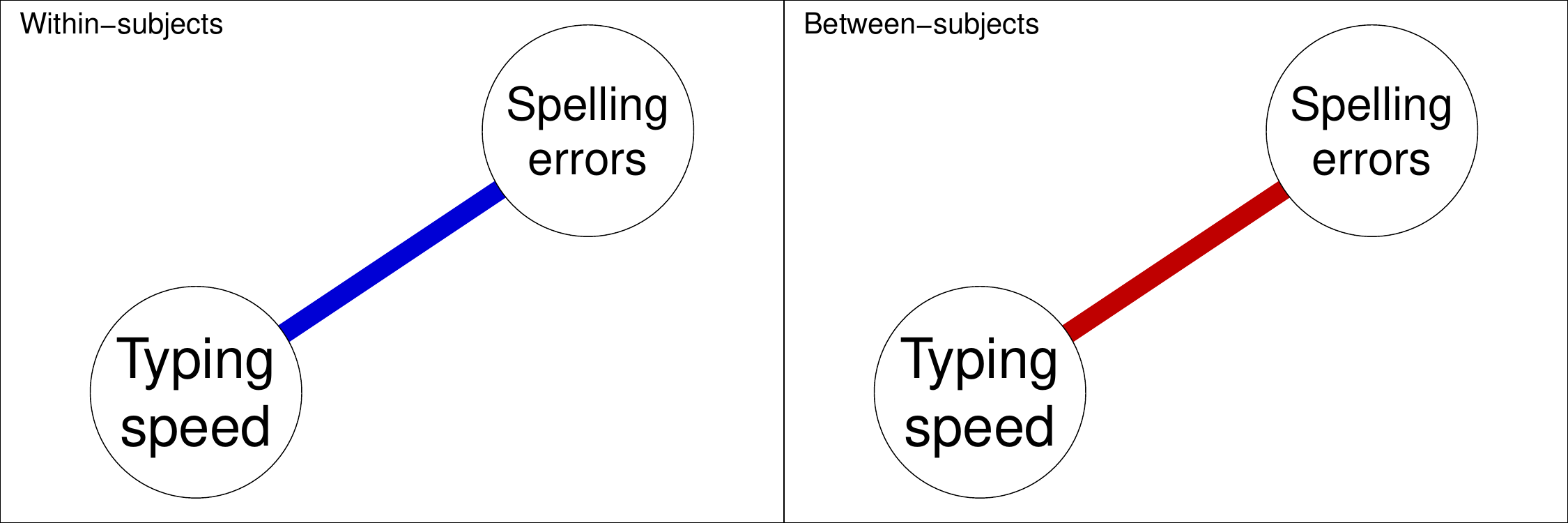}
        \caption{Example based on \protect\citeA{hamaker2012researchers}.}
\end{subfigure} \\
\begin{subfigure}{1\linewidth}
\centering
	  	\includegraphics[width=1\linewidth,page=2]{Figure3.pdf}
        \caption{Example based on \protect\citeA{hoffman2015longitudinal,EllenTalk}.}
\end{subfigure}
\caption{Two hypothetical examples of differing within- and between-subject networks. The networks on the left indicates the within-subject network, showing that personal deviations from the means predict each other at the same time point, and the networks on the right indicates the between-subjects network, showing how the \emph{means} of different subjects relate to one another.}
\label{dynamics:fig:mlVAR}
\end{figure}

\paragraph{Simpson's paradox} \citeA{hamaker2012researchers} described an example of how within- and between-subject effects can strongly differ from each other. Suppose we let people write several texts, and we measure the number of spelling errors they make and the number of words per minute they type (typing speed). We would expect to see the seemingly paradoxical network structures shown in Figure~\ref{dynamics:fig:mlVAR}, Panel~(a). We would expect a positive relationship in the within-subjects network (e.g., typing faster than your average leads to making more errors). Conversely, we would expect a negative relationship in the between-subject network (e.g., people who type fast, on average, generally make fewer spelling errors). This is because people who type fast, on average, are likely to be more skilled in writing (e.g., a courtroom stenographer) and are less prone to make a lot of spelling errors, compared to someone who types infrequently. Panel~(b) of Figure~\ref{dynamics:fig:mlVAR} shows another example in which the structures might differ (\citeNP{hoffman2015longitudinal}; provided by \citeNP{EllenTalk}). These network structures show that when people exert more physical activity than their average they likely experience an elevated heart rate, while people who on average are often physically active likely have a lower average heart rate. Such a different effect depending on the level of analysis is well known in the statistical literature as Simpson's paradox \cite{simpson1951interpretation}.

\paragraph{Interventionist accounts of causation} The different ways of thinking about the effects of manipulations in time-series models can be organized in terms recently developed from interventionist accounts of causation \cite{woodward2005making}. According to Woodward, causation is fleshed out in terms of interventions: $X$ is a cause of $Y$ if an intervention (natural or experimental) on $X$ leads to a change in $Y$. Statistically, the interventionist account is compatible with, for example, Pearl's \citeyear{pearl2000causality} semantics in terms of a ``do-operator.'' Here, an intervention on $X$ is represented as $\mathrm{Do}\left(X = x \right)$, and the causal effect on $Y$ is formally expressed as $\mathcal{E}\left(Y \mid \mathrm{Do}\left(X = x \right) \right)$. Pearl distinguished this from the classical statistical association, in which no intervention is present, and we get the ordinary regression $\mathcal{E}\left(Y \mid \mathrm{See}\left(X = x \right) \right)$. This notation is useful here, because it can be used to show how different kinds of causal manipulations, each at the intraindividual level, can produce a signal in either the between-subjects or the within-subjects network. 

Cashing out causal effects in terms of interventions is useful for understanding the intervention $\mathrm{Do}\left(X = x \right)$. We can think of this in terms of a random shock to the system, which sets $X$ to value $x$ at a particular time point and evaluates the effect on another variable $Y$ shortly afterwards. If we want to gauge this type of causal relationship, we might look at the within-subjects VAR model. Consider Hamaker's \citeyear{hamaker2012researchers} example regarding typing errors: If a researcher forced a person to type very fast, that researcher would need to evaluate the within-subject data, which would show a positive association between typing speed and the number of errors. In this example, between-subjects data would be misleading because individual differences would probably yield a negative correlation between speed and accuracy---faster typists are more likely to make less errors.

\paragraph{Interventions at the mean level} However, we can also think of a manipulation that sets $X$ to value $x$ in a different way, for instance, by inducing a long-term change in the system that leads it to converge on $X = x$ in expectation. To evaluate the effect of this type of intervention, it is important to consider the behavior of the system as it relates to the changes of the intercept of $X$. When analyzing time-series data gathered in a relatively short time-span, the within-subjects VAR network as discussed here cannot represent the relevant effects, because it assumes stationarity. However, such effects will be visible in the between-subjects network, which may thus contain important clues to the behavior of the system under potential changes in the intercept of one variable. In terms of Panel~(b) of Figure~\ref{dynamics:fig:mlVAR}, if we are interested in the effect of changing someone structurally---reducing the heart rate of a person on average---our preferred source of hypothesis generation would likely stem from the between-subjects model, as the corresponding within-subject model using the methods described in this paper only models deviations from the stationary mean. Such hypotheses could then be further investigated by using experimental design or lengthier longitudinal data analysis.

 Many such examples can be envisioned, especially in the field of psychopathology. For instance, short-term deviations from the mean in abusing a substance might not immediately develop tolerance or lead to one suffering from work or life inferences, but a subject who abuses a substance on average over a long time period might develop these problems (example based on variables used by \citeNP{rhemtulla2016network}). A between-subjects network could similarly show that loneliness mediates the effect of losing a spouse on depressive symptoms \cite{fried2015} or highlight the possible effects of childhood trauma and urbanization on psychotic symptoms \cite{isvoranua, isvoranu}---both cases in which within-subjects networks based on short-term deviations from the average seem less applicable. This analysis is important because it shows that, even though relevant causal interventions in psychology will typically operate at the intra-individual level, evidence for the effect of such interventions may arise at either the within- or the between-subjects level depending on the nature of the intervention.

\section{Empirical Examples}

\subsection{Reanalysis of M\~{o}ttus et al.\ (2017)}

We reanalyzed the data of \citeA{mottus2016within} to provide an empirical example of the multi\-level VAR methods described above. This data consists of two independent ESM samples, in which items tapping three of the five Five-Factor Model (neuroticism, extraversion, and conscientiousness; \citeNP{mccrae1992introduction}) domains were administered, as was an additional question that asked participants how much they had exercised since the preceding measurement occasion. Sample~1 consisted of 26 people providing $1{,}323$ observations in total, and Sample~2 consisted of 62 people providing a total of $2{,}193$ observations. Participants in Sample~1 answered questions three times per day, whereas participants in Sample~2 answered questions five times per day. In both samples, the minimum time between measurements was two hours. For more information about the samples and the specific questions asked, we refer readers to \citeA{mottus2016within}.

To obtain an easier and more interpretable example, we first only analyzed questions aimed as measuring the extraversion trait and the question measuring exercise. This lead to five variables of interest: questions pertaining to feeling outgoing, energetic, adventurous, or happy and the question measuring participants' exercise habits. We analyzed the data using the two-step multi\-level VAR procedure as described in detail in the Section~2 of the supplementary materials. We used the \emph{mlVAR} package, version 0.4, for the estimation of this model. Because the number of variables was small, we estimated the model using correlated temporal and contemporaneous random effects. We ran the model separately for both samples and computed the fixed effects for the temporal, contemporaneous, and between-subjects networks. Correlations of the edge weights indicated that all three networks showed high correspondence between the two samples (temporal network: 0.82, contemporaneous network: 0.94, between-subjects network: 0.70). Owing to the degree of replicability, we combined the two samples and estimated the model on the combined data.

\begin{figure*}
\centering
\includegraphics[width=1\linewidth]{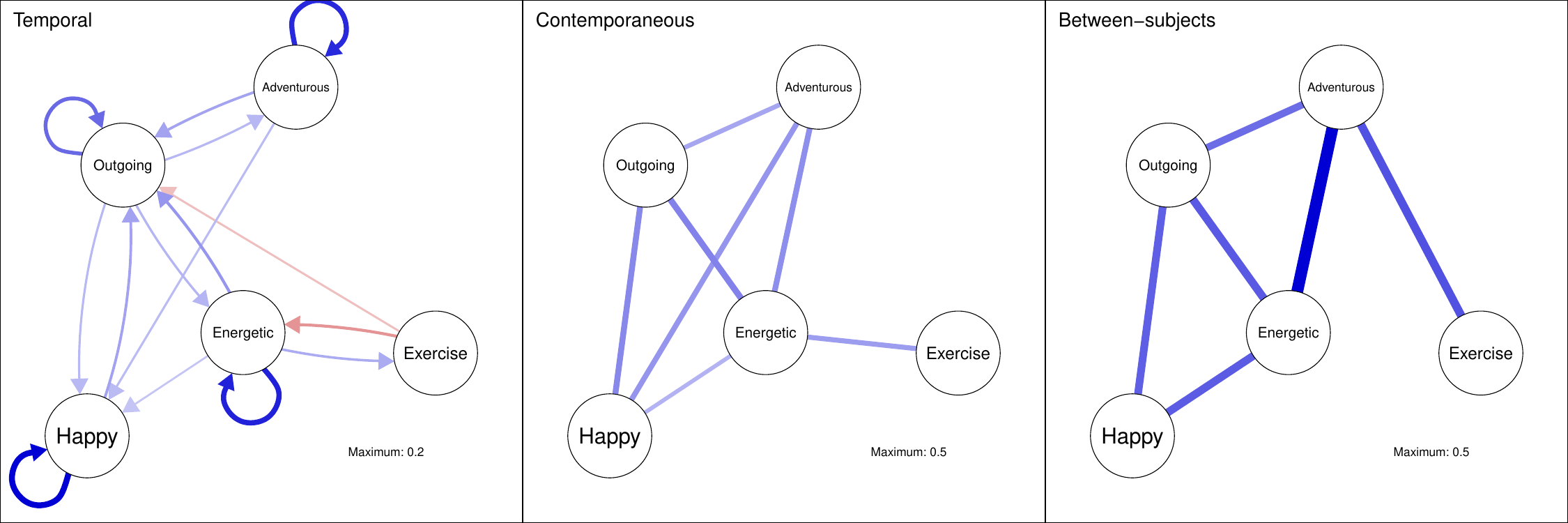}
\caption{The estimated fixed effects of the three network structures obtainable in multi\-level VAR. The model is based on ESM data of 88 people providing a total of $3{,}516$ observations. Due to differences in the scale of the networks, the temporal network was drawn with a different maximum value (i.e., the value indicating the strongest edge in the network) than the contemporaneous and between-subjects networks. Edges that were not significantly different from zero were removed from the networks.}
\label{dynamics:fig:example1fixed}
\end{figure*}

\begin{figure}
\centering
\includegraphics[width=1\linewidth,page=1,trim={0 0 3in 0}]{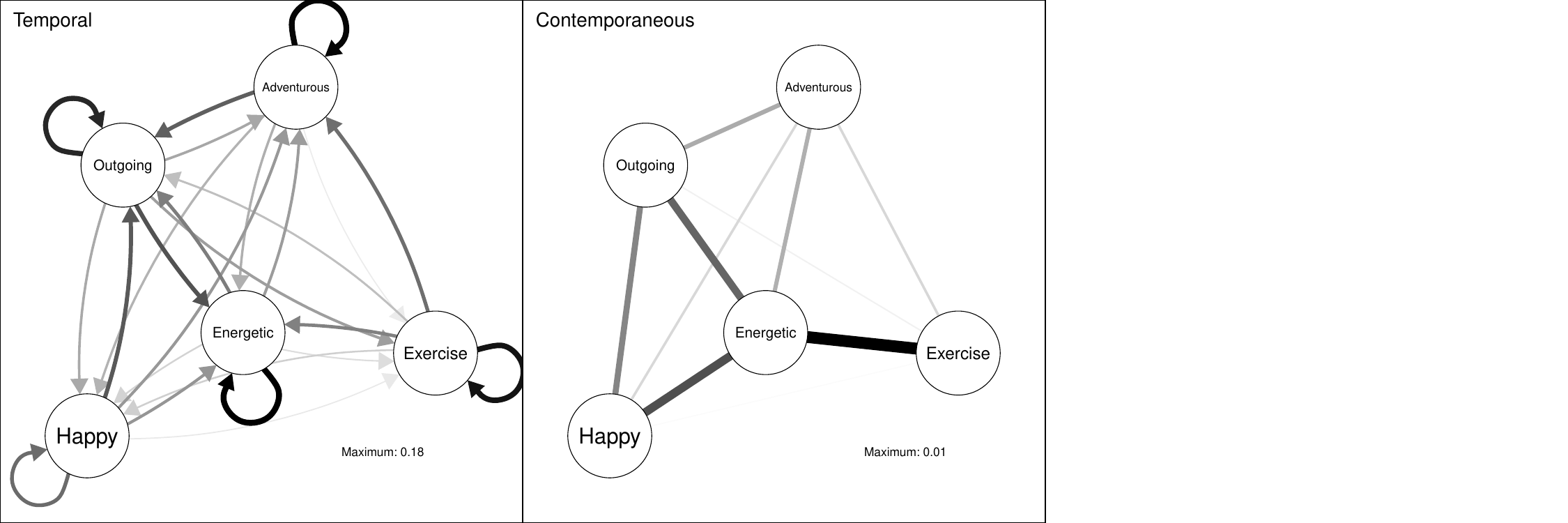}
\caption{The networks showing the standard deviation of random effects in the temporal and contemporaneous networks. Due to scale differences, networks were plotted using different maximum values.}
\label{dynamics:fig:example1random}
\end{figure}

\paragraph{Results} Figure~\ref{dynamics:fig:example1fixed} shows the estimated fixed effects of the temporal, contemporaneous, and between-subjects network. In these figures, only significant edges ($\alpha = 0.05$) are shown. In the contemporaneous and between-subjects networks, an edge was retained if one of the two regressions on which the partial correlation is based was significant (the so-called ``or'' rule; \citeNP{van2014new}). These results are in line with the hypothetical example shown in Figure~\ref{dynamics:fig:VAR}: People who exercised were more energetic while exercising and less energetic after exercising. In the between-subjects network, no relationship between exercising, feeling energetic, and feeling adventurous was found. The between-subjects network, however, showed a strong relationship between feeling adventurous and exercising: People who, on average, exercised more also felt, on average, more adventurous. This relationship was not present in the temporal network and much weaker in the contemporaneous network. Also noteworthy is that people were less outgoing after exercising. Figure~\ref{dynamics:fig:example1random} shows the standard deviation of the random effects in the temporal and contemporaneous networks. The largest individual differences in the temporal network were found in the auto-regressions, and the largest individual differences in the contemporaneous network were found in relationship between exercising and feeling energetic.

\begin{figure*}
\centering
\includegraphics[width=1\linewidth]{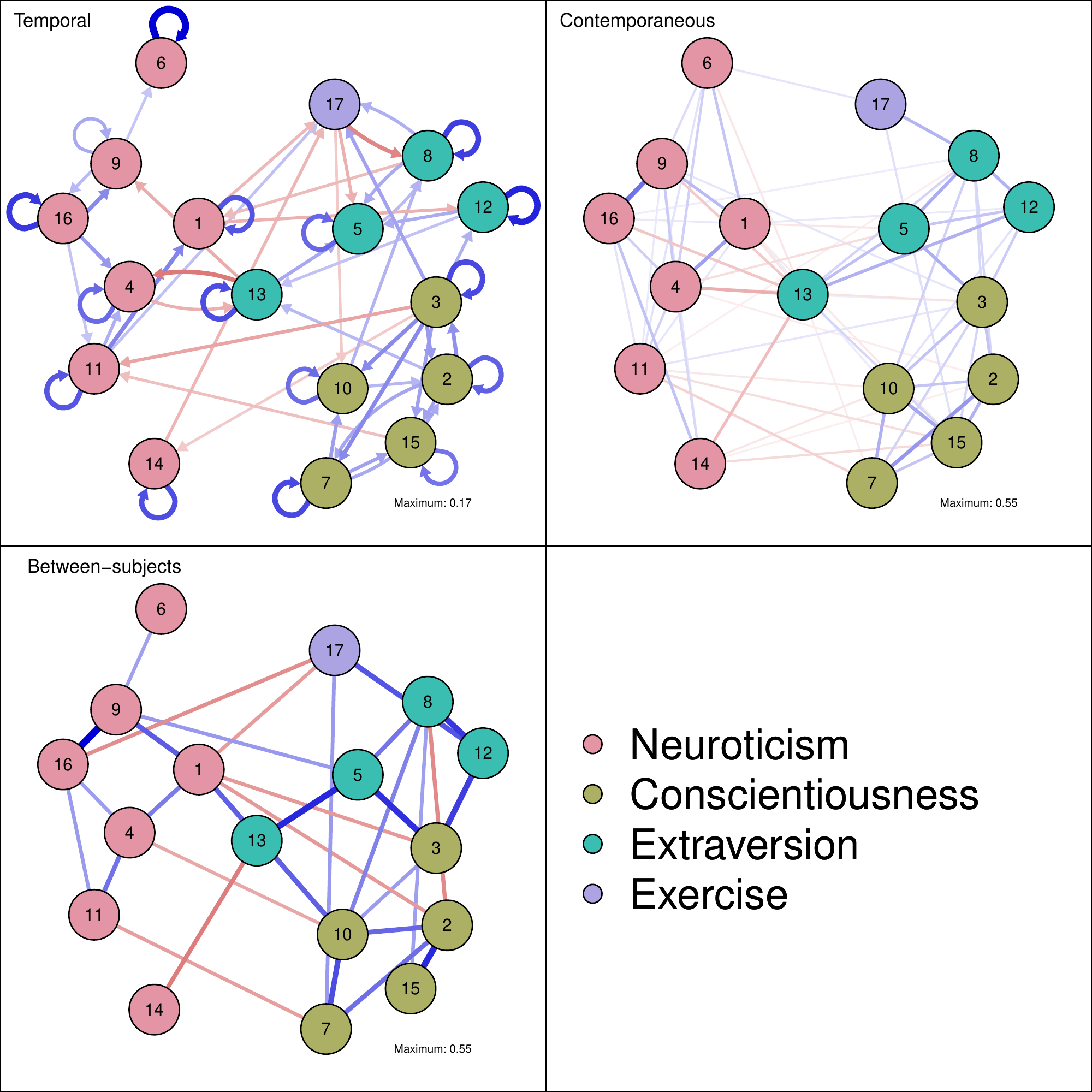}
\caption{The estimated fixed effects of the three network structures based on all 17 variables administered. Only significant edges are shown. Legend: 1 = ``Worried''; 2 = ``Organized''; 3 = ``Ambitious''; 4 = ``Depressed''; 5 = ``Outgoing''; 6 = ``Self-Conscious''; 7 = ``Self-Disciplined''; 8 = ``Energetic''; 9 = ``Frustrated''; 10 = ``Focused''; 11 = ``Guilty''; 12 = ``Adventurous''; 13 = ``Happy''; 14 = ``Control''; 15 = ``Achieved''; 16 = ``Angry''; 17 = ``Exercise.''}
\label{dynamics:fig:example2fixed}
\end{figure*}

In addition to using only the extraversion and exercise items, we also ran the model on all 17 administered items in the dataset. In this analysis, we used orthogonal random effects to estimate the model because correlated random effects cannot be estimated with such a large number of variables. Figure~\ref{dynamics:fig:example2fixed} shows the estimated fixed effects of the three network structures; it can be seen that indicators of the three traits tend to cluster together in all three networks. Regarding the node exercise, we found the same relationships between exercise, energetic, and adventurous (also found in the previous example) in the larger networks. Furthermore, we noted that exercising was connected to feeling angry in the between-subjects network but not in the other networks. Finally, there was a between-subjects connection between exercising and feeling self-disciplined: People who, on average, exercised more also felt, on average, more self-disciplined.

\subsection{Reanalysis of Bringmann et al.\ (2013)}

To showcase additional information that can be obtained using the GGM model, we reanalyzed the dataset used and made publicly available by \citeA{bringmann2013network}, which has been collected by \citeA{geschwind2011mindfulness}. This dataset contains ESM measures of 129 participants, which was collected in two periods over 6 days each: a baseline period and a posttreatment period (mindfulness treatment and a control group). Participants answered $60$ measurements per period. Similar to Figure~1 of \citeA{bringmann2013network}, we analyzed only the baseline dataset on the six items selected by \citeA{bringmann2013network}. We estimated the networks using three modeling frameworks discussed in Table~\ref{VARtable}. First, we analyze data using multi\-level Bayesian estimation using Mplus version 8 (model generated using the \emph{mlVAR} package). We estimated correlated random effects for the temporal effects but only fixed effects for the contemporaneous effects (making these random led to slow convergence). The model was estimated using three chains that ran until convergence. Nights were handled by adding a row of missing values between consecutive days. Second, we analyzed the data using two-step multi\-level VAR estimation as implemented in the \emph{mlVAR} package, using an ``and''-rule and estimating correlated random temporal and contemporaneous effects. Finally, we estimated the data using pooled and individual LASSO estimation using the \emph{graphicalVAR} package, using $\gamma = 0.25$. In the final two analyses, we did not regress the first measurement of the day on the last measurement of the previous day, and removed all pairs of lagged and current variables that contained missing responses. The final sample size was $5{,}927$ observations. Edges were retained if they were significant at the $\alpha=0.05$ level, or if $0$ was not included in the $95\%$ credibility interval.

\begin{figure*}
\centering
\begin{subfigure}[b]{1\linewidth}
	  	\includegraphics[width=1\linewidth,page=1]{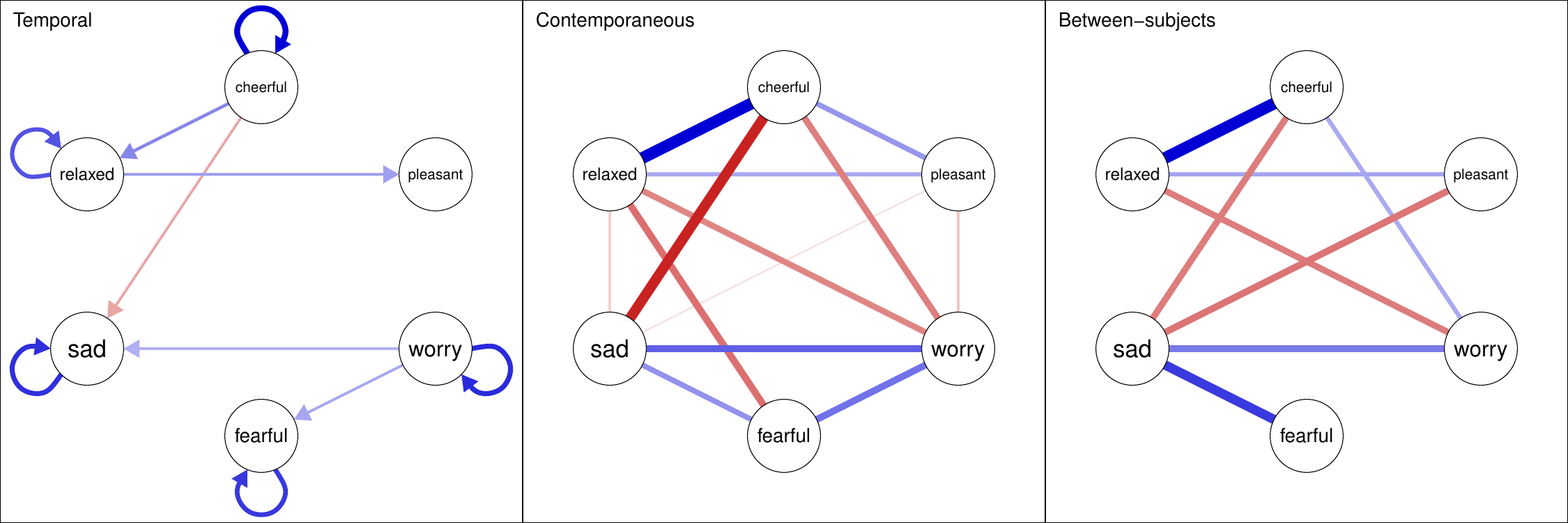}
        \caption{Fixed effect network structures estimated via multi\-level Bayesian estimation.}
\end{subfigure} \\
\begin{subfigure}[b]{1\linewidth}
	  	\includegraphics[width=1\linewidth,page=1]{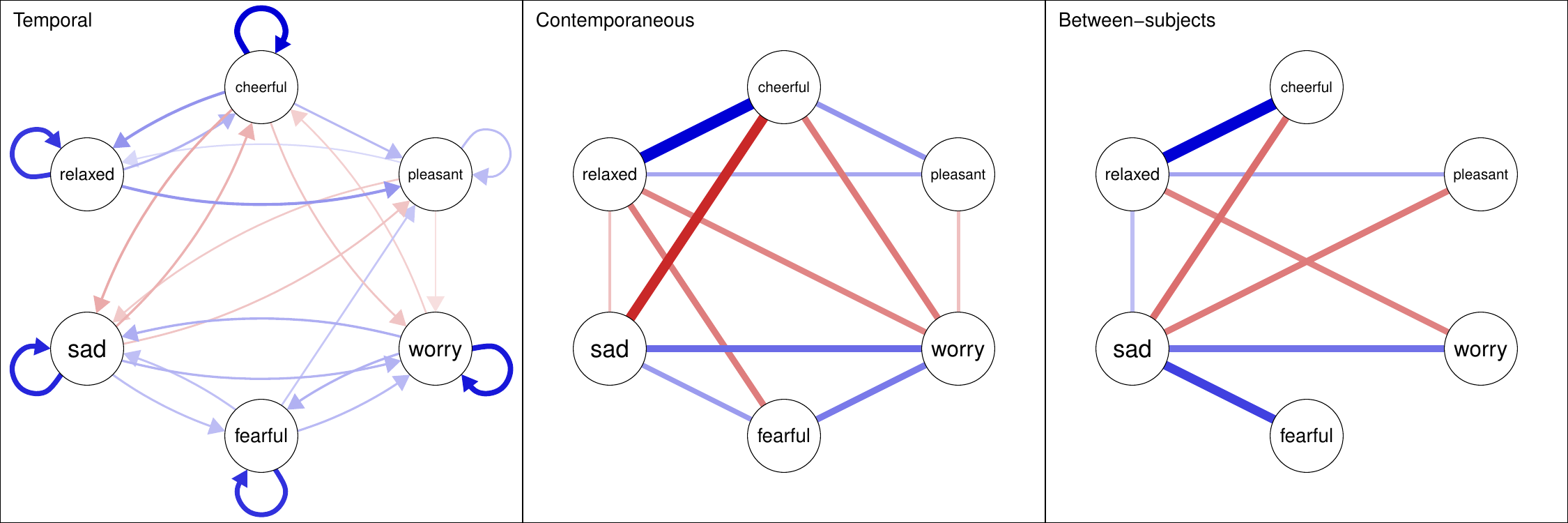}
        \caption{Fixed effect network structures estimated via two-step multi\-level estimation.}
\end{subfigure} \\
\begin{subfigure}[b]{1\linewidth}
	  	\includegraphics[width=1\linewidth,page=1]{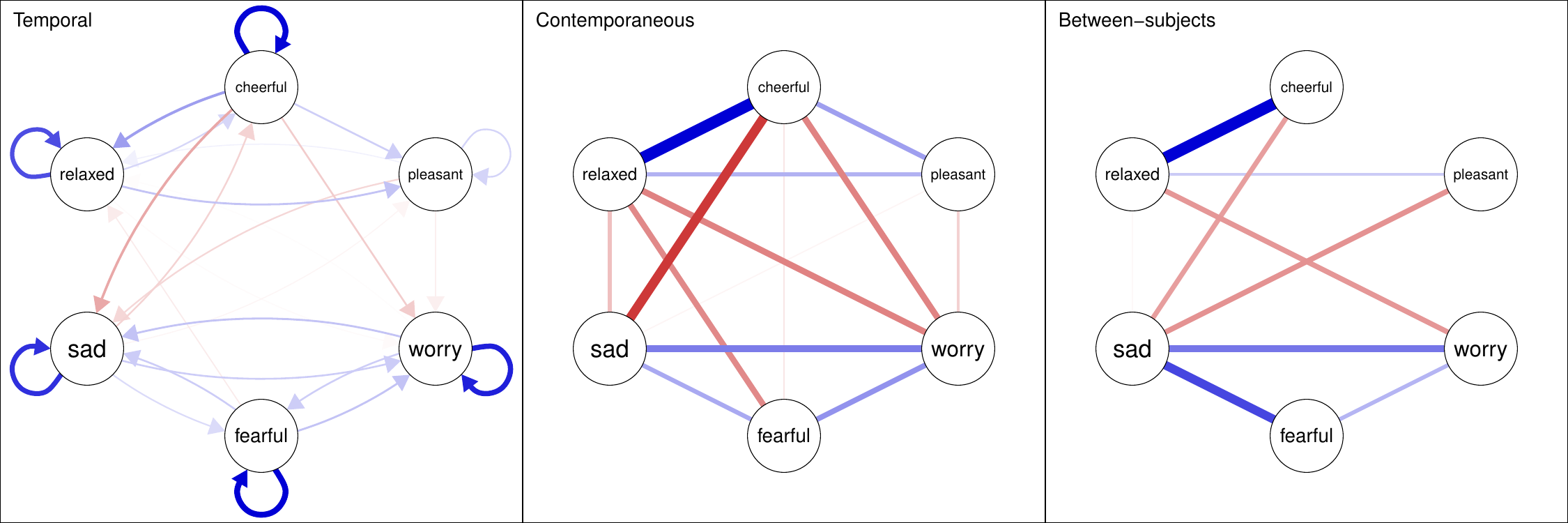}
        \caption{Fixed effect network structures estimated via pooled and individual LASSO estmiation}
\end{subfigure} \\
\caption{Reanalysis of the \protect\citeA{geschwind2011mindfulness} dataset used by \protect\citeA{bringmann2013network}.}
\label{dynamics:fig:bringmann}
\end{figure*}

\paragraph{Results} Figure~\ref{dynamics:fig:bringmann} shows the resulting network structures, and shows that all three methods are mostly aligned. Unsurprisingly, the temporal networks are very similar to those reported by \citeA{bringmann2013network}.\footnote{The networks differ because the estimation of temporal effects differs in that measures are within-subjects centered and subject means are included as Level 2 predictors.} Both the temporal and contemporaneous network are in line with what would be expected under a unidimensional auto-correlated latent variable model (many edges selected, low-rank structure, edges of expected sign) with the exception of the positive temporal edge from ``fearful'' to ``pleasant'' in the two-step multi\-level network (which was not selected by the other methods). Of note is that Bayesian multi\-level estimation resulted in a sparser temporal network. This difference in sparsity is possibly because the multivariate Bayesian multilevel approach more accurately represents the uncertainties in parameter estimation, while two-step multilevel VAR estimates the model piecewise and pooled LASSO estimation does not take the multilevel structure into account. Remarkable is the positive edge between ``sad'' and ``relaxed'' in the two-step multi\-level between-subjects network, which is based on two significant positive Level~2 regression coefficients ($\beta = 0.202, p = 0.046$ and $\beta = 0.151, p = 0.036$) where the estimated between-subjects correlation is strongly negative ($-0.53$). This edge is especially remarkable since both nodes are  strongly connected to other nodes in the network. The Bayesian multi\-level between-subjects network showed a similar positive edge between ``cheerful'' and ``worry''. This is noteworthy because under a unidimensional factor model, we would not expect partial correlation coefficients to switch sign from marginal correlation coefficients \cite{holland1986conditional,riet}. A possible way the partial correlation coefficient switches sign is if it has been conditioned on one or more common effects between the two variables of interest (in this case, potentially ``worry,'' ``pleasant,'' or ``fearful''). Of course, these effects must be interpreted with great care, especially given the high $p$-values; we did not control for multiple comparisons, and the same edges are not retained in the other methods. Still, it is noteworthy that if this edge is weak or nonexistent, the between-subjects structure is still not in line with a unidimensional factor model. In such a factor model, ``sad'' and ``relaxed''(which feature the most connections) would be expected to have a strong negative edge between them (a depression factor would lead to ``sad'' having a strong positive factor loading and ``relaxed'' having a strong negative factor loading).

\section{Discussion}

We discussed the Gaussian graphical model (GGM; \citeNP{lauritzen1996graphical}), an undirected network model of partial correlation coefficients, and discussed its utility in the analysis of psychological datasets. The GGM presents a promising exploratory data analysis tool that allows for different levels of interpretation: (1) Edges in the GGM can be interpreted without reliance on a causal interpretation and merely used to show which variables predict each other. (2) Causal effects between variables result in an edge, whereas the lack of a causal effect results in no edge, except in the presence of latent variables or a common effect. The GGM can, therefore, be seen as hypothesis generating structures that highlight potential causal pathways. (3) Undirected models can be used and interpreted as causal data-generating process and have been used as such in several fields of research. 

The GGM can readily be estimated on any dataset that contains multiple observations of the same variables (e.g., multiple people in cross-sectional data or multiple responses in time-series data). LASSO regularization methods perform especially well in estimating such a GGM structure. In temporally ordered data (e.g., $n = 1$ time series), the graphical VAR (GVAR; \citeNP{wild2010graphical}) model generalizes the GGM to incorporate temporal effects. We showed how two network structures can be obtained: a temporal network, which is a directed network of regression coefficients between lagged and current variables, and a contemporaneous network, which is a GGM describing the relationships that remain after controlling for temporal effects. In temporally ordered data of multiple subjects (e.g., $n >  1$ time series), the natural combination of cross-sectional and time-series data came by adding a third network structure: the between-subjects network, which is a GGM that describes relationships between the stationary means of subjects. We proposed two methods to estimate the three network structures: (1) two-step multi\-level estimation, which we implemented in the open source R package \emph{mlVAR}, and (2) pooled and individual VAR model estimations using LASSO regularization, which we implemented in the open source R package \emph{graphicalVAR}.

\subsection{Limitations and Challenges}

\paragraph{Multilevel estimation} The presented methods are not without problems and have several limitations. With regard to multi\-level estimation, first, multivariate estimation of the multi\-level VAR model is not yet feasible for larger datasets. As such, the proposed two-step multi\-level VAR combines univariate models. Doing so, however, means that not all parameters are in the same model. In addition, univariate models do not readily provide estimates of the contemporaneous networks, which must be estimated in a second step. Second, even when multivariate estimation is possible, it is still challenging to estimate a multi\-level model on contemporaneous networks due to the requirement of positive definite matrices. Third, when more than approximately eight variables are measured, estimating the multi\-level models with correlated random effects is no longer feasible in open source software. In this case, orthogonal random effects can be used, which induce a level of parsimony that may not be substantively plausible. Finally, even when orthogonal estimation is used, multi\-level analysis runs very slowly in models with more than $20$ variables. As such, multi\-level VAR analysis of high-dimensional datasets is not yet feasible. To this end, we discussed pooling within-subject centered data and estimating fixed-effects models using LASSO regularization \cite{abegaz2013sparse}. This performed on par with multi\-level estimation in higher sample sizes and allows researchers to scale up the analysis. However, individual network estimation using separate VAR models does not borrow information from other subjects and performs poorly in low sample sizes. Promising developments are new LASSO methods in which shrinkage from subject-specific parameters to their mean is attained through penalization rather than hierarchical modeling \cite{hastie2015statistical}. Future research should investigate the utility of such models in estimating individual network structures that might differ in structure but borrow information from other subjects in its estimation.

\paragraph{VAR modeling assumptions} These limitations on the estimation methods come with more limitations in the statistical models themselves. VAR modeling, especially, is not without problems and faces severe challenges \cite{hamaker2015modeling, hamakerPresent}. We made several assumptions that can be problematic. For instance, in characterizing the likelihood of time-series data, we need to assume that the conditional distribution of variables at time $t$  given time $t-1$ are the same for all $t$. That raises two distinct assumptions: (1) The difference in time between measurements are roughly equal, and (2) the parameters do not change over time. Equidistance in time is especially important for the interpretation of temporal networks. Promising work is being done in this area where VAR networks can be estimated on nonequidistant datasets \cite{driverCont, oravecz2009hierarchical,oud2000continuous}. The assumption of stationarity is needed to estimate structures when data are limited but might not be tenable especially in longer time series \cite{rovine2006multilevel}. Promising time-varying estimation procedures are being developed \cite{bringmann2016changing, mgm}, but are not yet extended to the GVAR framework. Furthermore, the interpretation of temporal coefficients when represented as a network is not without discussion, and several different methods for standardization exist \cite{bulteel,schuurman2016compare}.\footnote{We standardized every dataset before analyzing and used the standardization of \citeA{wild2010graphical} for temporal networks in $n = 1$ and pooled temporal networks. GGMs are readily standardized by using partial correlation coefficients (Equation~\eqref{eq:parcor}), which have been used in all GGMs shown in this paper.}

\paragraph{Normality} Another particularly important assumption made in this paper is that of multivariate normality. Indeed, Equation~\eqref{yisnormal} makes this assumption and all other equations follow from this. The assumption of normality is not without problems \cite{terluin2016differences}. However, it is not always straightforward to deal with these issues, because violations of normality may arise for many different reasons. When data are not normally distributed, then they cannot be represented properly using only the means vector and variance--covariance matrix. As a result, the GGM does not properly characterize the joint likelihood function. When data are measured on a different scale \cite{stevens1946theory}, a different graphical model can be used, such as the Ising model for binary data \cite{netpsych, van2014new} or a mixed graphical model for categorical and Poisson-distributed variables as well as binary and Gaussian variables \cite{jonas2}. Such models have yet to be extended to time-series analysis, especially in separating temporal and contemporaneous effects as the GVAR model does. When data are continuous but not normal, multiple reasons can (again) contribute to this. When the underlying process is normal but the measured variables are on a transformed scale, transforming data back to normal should offer a solution \cite{liu2009nonparanormal}, but when the process itself is nonnormal, such as skewed residuals, the entire modeling framework does not correctly capture the likelihood. Finally, multivariate normality assumes all relationships between variables are linear. When this is not the case, the GGM and VAR model (which fit linear effects) will not properly describe the data. We encourage future researchers to focus on the problem of normality and to develop new methods of overcoming these challenges.

\paragraph{Interpretation} Finally, it should be noted that when taking a causal interpretation of edges, all methods discussed in this paper are exploratory in nature and can only generate hypotheses---they do not confirm causal relations. The analyses showcased in this paper can also be used without relying on a causal interpretation and allow researchers to obtain insights into the predictive relationships present in the data---regardless of theory with respect to the data-generating model. Under the assumptions of multivariate normality, stationarity, and the Lag-$1$ factorization, the networks show how variables predict each other over time (temporal network), within time (contemporaneous network), and on average (between-subjects network). Furthermore, during the thresholding of edges in the multi\-level analyses, we did not apply a correction for multiple testing by default. We deliberately chose this because our aim was to present exploratory hypothesis-generating structures, and not correcting for multiple testing yields greater sensitivity.

\subsection{Conclusion}

	This paper provides a methodological overview of how the GGM can be used in various different kinds of psychological data. The GGM can be used to map out unique variance in cross-sectional data or at the contemporaneous and between-subjects levels of time-series analysis. We contrasted this method to exploratory estimation of causal models. While losing information on the direction of effect, estimating GGMs offers an attractive alternative in that these models are uniquely identified, well parameterized, closely related to causal models and also offer exploratory insight on predictive effects between observed variables. When the aim is to discover psychological dynamics, the GGM can be used as a hypothesis generating technique inspiring future research or therapy directions \cite{contempraneous,KroezeR}. For example, an effect found in a cross-sectional analysis could inspire a time-series study, a contemporaneous effect could inspire a shorter time-lag time-series study and a between-subjects effect could inspire lengthy longitudinal studies. All network structures may inspire experimental design, or to gather a mixture of observational and experimental data \cite{magliacane2017causal}. The GGM thus provides a powerful addition to the exploratory toolbox in behavioral research.

\section{Acknowledgements}

We would like to thank Laura Bringmann, No\'{e}mi Schuurman, Ois\'{i}n Ryan and Ellen Hamaker for helpful tips and invigorating discussions, and Katharina Jorgensen for valuable comments on earlier versions of this paper. 


\clearpage
\beginsupplement

\section{Appendix A. Glossary of terms}

\label{glossary}

{\footnotesize
\begin{center}
		\begin{tabular}{l p{10cm}}
			\toprule
			Term & Explanation  \\
			\midrule
			Undirected network & A network model in which nodes are connected by edges (also termed links) without arrowheads. \\
			Directed network & A network model in which nodes are connected by edges with arrowheads, assumed to display causal effects or temporal prediction. \\
Gaussian graphical model & An undirected network model in which observed variables are represented with nodes. Nodes are connected with an edge if two variables are not independent after conditioning on all other observed variables. Edges are parametrized by using partial correlation coefficients. \\
	Causal model & A causal model of observed and unobserved variables that is assumed to generate the data. \\
	Directed acyclic graph & A directed network in which one node does not eventually point to itself. \\
	 Within-subjects network & A network model explaining within-subject (co)variation from the stationary mean. \\
	 Between-subjects network & A network model explaining (co)variation between stationary means of different persons. \\
	 Cross-sectional network & A network model estimated on cross-sectional data. Can be shown to be a blend of the within-subjects and between-subjects networks. Can be interpreted as representative of within-subjects or between-subjects network based on the way in which data is gathered. \\
	 Vector auto-regression (VAR) & Multivariate regression of a set of variables on previous realizations of that set of variables. \\
	 Temporal network & A within-subject network model of effects between different measurement occasions, showing temporal prediction or potential causal pathways. \\
	 Contemporaneous network & A within-subject undirected network model of effects between variables in the same measurement occasion, after taking temporal effects into account. \\
			\bottomrule
		\end{tabular}
\end{center}}

\clearpage

\section{Supplementary 1: Notation}

Throughout the paper we employ the following notation. Roman letters indicate observed variables, and Greek letters indicate parameters or latent variables. Nonboldface letters indicate a single value. An uppercase nonboldface letter indicates a random variable, and a lowercase nonboldface letter indicates a realization. We use $t$ to denote measurement occasion and $T$ to denote a random measurement occasion,\footnote{Mostly we assume measurements are nested in subjects, and two subjects might have a different number of measurement occasions. As such, $t=1$ for subject $p=1$ might not correspond to $t=1$ for subject $p=2$.} $i$ ($i \in \{1, 2, \ldots, m\}$) to denote item administered, and $p$ ($p \in \{1, 2, \ldots, n \}$) to denote a subject and $P$ to denote a random subject. We will use lowercase boldface letters to denote column vectors and uppercase boldface vectors to denote matrices. Subscripts will denote if these are random or fixed. For example, $\pmb{B}_p$ will denote a fixed matrix for subject $p$, and $\pmb{B}_P$ will denote the matrix of random subject $P$ (which has a distribution). 

Because we are interested in finding dynamics between items, we use vector $\pmb{y}$ to denote the set of all items.\footnote{If researchers are interested in dyadic interactions \cite{ferrer2016exploratory}, for example, then a dyadic pair can be seen as a ``subject,'' and items can be the item responses from both subjects.} For the observed variables, we will use consistent subscripts (measurement, subject) to denote which items are contained in the vector. For example, $\pmb{y}_{[t,p]}$ denotes all responses of subject $p$ at time point $t$, and $\pmb{y}_{[T,p]}$ denotes all responses of subject $p$ at a random time point $T$. A set in this subscript indicates multiple responses. For example, we will use $\pmb{y}^\top_{[\{t-1,t\},p]} = \begin{bmatrix} \pmb{y}^\top_{[t-1,p]} & \pmb{y}^\top_{[t,p]} \end{bmatrix}$ to denote a set of lagged and current responses from subject $p$ around time point $t$. If only one observation or subject is measured, we will drop the square brackets (e.g., $\pmb{y}_{P} = \pmb{y}_{[1,P]}$ indicates the cross-sectional response pattern of a random subject). When it is unclear if the set of items corresponds to a random person or a random measurement occasion, we refer to $C$ as a random case, with $c$ as a particular case, and subset the data either as $\pmb{y}_{C}$ to describe a random response pattern or $\pmb{y}_{c}$ to describe a realization---in cross-sectional data $\pmb{y}_{C} = \pmb{y}_{P}$ and in $N=1$ time-series data $\pmb{y}_{C} = \pmb{y}_{T}$. $C$ could also indicate a set of multiple responses. Other subscripts denote subsets of a vector or matrix, with notation $-(\ldots)$ indicating the subset of everything except $\{\ldots\}$.

\section{Supplementary 2: Two-step multi-level VAR}

\label{twostepmlVAR}

In this appendix, we will outline two-step multi-level VAR, which we propose as a methodology to estimate the GVAR model using multi-level estimation. This method builds on the work of \citeA{bringmann2013network}, and extends their proposed algorithm by including between-subject effects \cite{hamaker2014center} and estimating the contemporaneous network by performing a second multi-level estimation on the residuals of the temporal model (the second ``step''). To reiterate the paper, the model to estimate is:
\begin{equation*}
\pmb{y}_{[T,p]}  \mid \pmb{y}_{[T-1,p]} = \pmb{y}_{[t-1,p]}  \sim N\left( \pmb{\mu}_p +  \pmb{B}_{p} \left(\pmb{y}_{[t-1,p]} -  \pmb{\mu}_p\right), \pmb{\Theta}_p   \right).
\end{equation*}
In particular, we are interested in estimating between-subjects network $\pmb{K}^{(\pmb{\Omega})} = \pmb{\Omega}^{-1} = \mathrm{Var}(\pmb{\mu}_{P})^{-1}$ and the (distributions of) temporal networks $\pmb{B}_{p}$ and contemporaneous networks $\pmb{K}^{(\pmb{\Theta})}_{p} = \pmb{\Theta}^{-1}_{p}$.

\subsection{Multi-level modeling}

The fixed effects and random effect variances and covariances can be estimated by estimating a VAR model for every subject, pooling the parameter estimates, and computing the mean (fixed effects) and variance--covariance matrix (random effects distribution). This estimation, however, is separate for every subject. To combine all observations in a single model, we can assign distributions over the parameters; in which case, we make use of multilevel modeling. Assigning distributions has two main benefits. First, instead of having a single parameter per subject, we now only need to estimate the parameters of the distribution. For example, when we model observations from 100 subjects, instead of estimating each parameter 100 times, we now only need to estimate its mean and variance. Second, the multilevel structure acts as a prior distribution in Bayesian estimation procedures---in case we wish to obtain person-specific parameter estimates post hoc. In particular, multilevel modeling leads to \emph{shrinkage}; parameter values that are very different from the fixed effects are likely to be estimated closer to the fixed effect in multilevel modeling than when using a separate model for every subject. For example, if we estimate a certain temporal regression in five people and find the values $1.1$, $0.9$, $0.7$, $1.3$, and $10$, it is likely that the fifth statistic, $10$, is an outlier. Ideally, we would estimate this value to be closer to the other values.

\citeA{bringmann2013network} proposed a sequential univariate method for estimating temporal VAR models. Because the joint conditional distribution of $\pmb{y}_{[T,p]} \mid \pmb{y}_{[T-1,p]} = \pmb{y}_{[t-1,p]}$ is normal, it follows that the marginal distribution of every variable is univariate normal and can be obtained by dropping all other parameters from the distribution:
\begin{equation*}
 y_{[T,p,i]} \mid \pmb{y}_{[T-1,p]} = \pmb{y}_{[t-1,p]} \sim N\left( \mu_{[p,i]} +  \pmb{\beta}_{[p,i]} \left(\pmb{y}_{[t-1,p]} - \pmb{\mu}_{p}\right), \theta_{[p,i]}   \right),
\end{equation*}
in which $ y_{[T,p,i]}$ denotes the $i$th element of $\pmb{y}_{[T,p]}$, $\pmb{\beta}_{[p,i]}$ indicates the row vector of the $i$th row of $\pmb{B}_{p}$, and $\theta_{[p,i]}$ denotes the $i$th diagonal element of $\pmb{\Theta}_p$. When drawn as a temporal network, the edges point to node $i$. Many software packages do not allow the estimation of $\pmb{\mu}_{p}$ as described above. In this case, the sample means of every subject,  $\bar{\pmb{y}}_{p}$, can be taken as a substitute for $\pmb{\mu}_{p}$ \cite{hamaker2014center}. The model then becomes a univariate multilevel regression model with within-subject centered predictors, estimable by functions such as the \textVerb{lmer} in \textVerb{lme4} \cite{lme4}. The Level 1 model becomes
\begin{align}
\label{dynamics:univariatemlVAR}
  y_{[t,p,i]} &= \mu_{[p,i]} +  \pmb{\beta}_{[p,i]} \left(\pmb{y}_{[t-1,p]} - \bar{\pmb{y}}_{p}\right) + \varepsilon_{[t,p,i]} \\
\varepsilon_{[T,p,i]}  &\sim N(0,\theta_{[p,i]} ),
\end{align}
and the Level 2 model becomes
\begin{equation*}
\begin{bmatrix}
\mu_{[P,i]} \\
\pmb{\beta}_{[P,i]}
\end{bmatrix} \sim N\left(
\begin{bmatrix}
0 \\
\pmb{\beta}_{*i}
\end{bmatrix}
,
\begin{bmatrix}
\omega_{\mu_i} & \pmb{\omega}^{(\pmb{\beta}_i\mu_i)\top} \\
\pmb{\omega}^{(\pmb{\beta}_i\mu_i)} & \pmb{\Omega}^{(\pmb{\beta}_i)}
\end{bmatrix}
\right).
\end{equation*}
Estimation of such univariate models requires integrating over a simpler integral than estimation of multivariate models. As a result, sequential estimation using univariate models have been used in estimating multilevel VAR models \cite{bringmann2013network}. A downside, however, is that not all parameters are included in the model.  In particular, correlations between means (between-subject effects) and between contemporaneous covariances are not retained, as well as the correlations between temporal edges pointing to different nodes. A second downside is that estimating correlated random effects does not work well for models with many predictors. In particular, \textVerb{lmer} becomes very slow with approximately more than eight predictors. As such, networks with more than eight nodes are hard to estimate. To estimate larger networks (e.g., 20 nodes), we can choose to estimate uncorrelated random effects, which we term \emph{orthogonal estimation}.

\subsection{Extending multi-level VAR: two-step multi-level VAR}

The methodology of \citeA{bringmann2013network} does not estimate contemporaneous or between-subjects networks. Therefore, we propose extensions to the algorithm to estimate these networks. We propose a two-step method. Step 1 follows the procedure of \cite{bringmann2013network} with the addition that between-subject effects are included \cite{hamaker2014center}. This leads to estimates of the temporal and between-subjects networks. The second step involves taking the residuals of step 1 in order to obtain contemporaneous networks.

\paragraph{Step 1: Temporal and between-subjects networks} To obtain estimates of between-subject effects, the sample means of every subject, $\bar{\pmb{y}}_{p}$ in Equation~\eqref{dynamics:univariatemlVAR}, can be included as predictors at the subject level (except for the mean of the dependent variable; \citeNP{hamaker2014center,hoffman2009persons,curran2011disaggregation}). With this extension, the Level 2 model for the person-specific mean of the $i$th variable now becomes
\begin{equation}
\label{dynamics:eq:meannodewise}
\mu_{[p,i]} = \pmb{\beta}^{(\mu)}_{i} \bar{\pmb{y}}_{[p,-(i)]} + \varepsilon_{[p,i]}^{(\mu)},
\end{equation}
in which we use $\pmb{\beta}^{(\mu)}_{i}$ to denote the $i$th row  (without the diagonal element $i$) of an $m \times m$ matrix $\pmb{B}^{(\mu)}$, and $\bar{\pmb{y}}_{[p,-(i)]}$ denotes the vector $\bar{\pmb{y}}_{p}$ without the $i$-th element. Because $\bar{y}_{[p,i]}$ is itself an estimate of $\mu_{[p,i]}$, Equation~\eqref{dynamics:eq:meannodewise} takes the form of a multiple regression model. As such, these estimates can be used to estimate a GGM between the means \cite{lauritzen1996graphical,meinshausen2006high}---the between-subjects network:
\[
\pmb{K}^{(\pmb{\mu})} \approx  \pmb{D}^{(\pmb{\mu})} \left(\pmb{I} - \pmb{B}^{(\mu)}\right),
\]
with $d^{(\mu)}_{ii} = 1/\mathrm{Var}(\varepsilon_{[P,i]}^{(\mu)})$.
Due to the estimation in a multilevel framework, the resulting matrix will not be perfectly symmetric and must be made symmetric by averaging lower and upper triangular elements. Thus, each edge (i.e., partial correlation) in the between-subjects network is estimated by standardizing and averaging two regression parameters: the parameter denoting how well mean $A$ predicts mean $B$ and the regression parameter denoting how well mean $B$ predicts mean $A$. Obtaining the between-subjects effects using regression coefficients rather than correlating the estimated means leads to standard errors that can be used to select significant edges. 

\paragraph{Step 2: Contemporaneous networks} An estimate for contemporaneous networks can be obtained in a second step by investigating the residuals of the multilevel model that estimate the temporal and between-subject effects. These residuals can be used to run multilevel models that predict the residuals of one variable from the residuals of other variables at the same time point. Let $\hat{\varepsilon}_{[t,p,i]}$ denote the estimated residual of variable $i$ at time point $t$ of person $p$, and let $\hat{\pmb{\varepsilon}}_{[t,p,-(i)]}$ denote the vector of residuals of all other variables at this time point. The Level 1 model then becomes
\begin{equation}
\label{dynamics:eq:residnodewise}
\hat{\varepsilon}_{[t,p,i]} = \pmb{\beta}^{(\pmb{\Theta})}_{[p,i]} \hat{\pmb{\varepsilon}}_{[t,p,-(i)]} + \varepsilon^{(\pmb{\Theta})}_{[t,p,i]},
\end{equation}
in which $\pmb{\beta}^{(\pmb{\Theta})}_{[p,i]}$ represents the $i$-th row (without the diagonal element $i$) of an $m \times m$ matrix, $\pmb{B}^{(\pmb{\Theta})}_{p}$, and $\varepsilon^{(\pmb{\Theta})}_{[t,p,i]}$ represents a residual. In the Level 2 model, we again assign a multivariate normal distribution to parameters in $\pmb{\beta}^{(\pmb{\Theta})}_{i}$. It can be seen that Equation~\eqref{dynamics:eq:residnodewise} also takes the form of a multiple regression model. Thus, this model can again be seen as the node-wise GGM estimation procedure:
\[
\pmb{K}^{(\pmb{\Theta})}_p \approx  \pmb{D}^{(\pmb{\Theta})}_p \left(\pmb{I} - \pmb{B}^{(\pmb{\Theta})}_p\right),
\]
with $d^{(\pmb{\Theta})}_{[p,i]} = 1/\mathrm{Var}(\varepsilon_{[T,p,i]}^{(\pmb{\Theta})})$. Again the matrices need to be made symmetric by averaging upper and lower triangle elements. By using univariate multi-level regressions, rather than simply correlating the residuals, we can impose multi\-level structure on the partial correlations in order to estimate fixed and random effects.\footnote{ Estimating correlated random effects for regression coefficients is straightforward while estimating correlated random effects on covariances or correlations is not due to a requirement that these must add to a positive definite variance--covariance matrix; fixed effects covariances plus multivariate normally distributed random effects may lead to intra-individual variance--covariances that are not positive-definite.}  Fixed effects can be obtained by using the fixed effects matrices instead in the expression above. As with the temporal network, orthogonal estimation can be used when the number of variables is large (i.e., larger than approximately eight).

\paragraph{Thresholding} After estimating network structures, researchers may be interested in removing edges that may be spurious and due to sampling error. By setting edge weights to zero, effectively removing edges from a network, a sparse network is obtained that is more easily interpretable. One method of doing so is by removing all edges that are not significantly different from zero. For fixed effects, multilevel software returns standard errors and $p$-values, allowing for this thresholding. For the temporal networks, each edge is represented by one parameter and thus by one $p$-value. The contemporaneous and between-subjects networks, however, are a function of two parameters that are standardized and averaged: a regression parameter for the multiple regression model of the first node and a regression parameter for the multiple regression model of the second node. As such, for every edge, two $p$-values are obtained. We can choose to retain edges of which at least one of the two $p$-values is significant, termed the ``or'' rule, or we can choose to retain edges in which both $p$-values are significant, termed the ``and'' rule \cite{barber2015high}.

\paragraph{Summary} In sum, the above described two-step estimation method proposes to estimate a multilevel model per variable, using within-person centered lagged variables as within-subject predictors and the sample means as between-subject predictors. These models can be used to obtain estimates for the temporal network and between-subjects network. In a second step, the contemporaneous networks can be estimated by estimating a second multilevel on the residuals of the first multilevel model. The \emph{mlVAR} R package  implements these methods \cite{mlvar}. In this package, temporal coefficients can be estimated as being ``unique'' per subject (unique VAR models per subject), ``correlated'' (estimating correlations between temporal effects), ``orthogonal'' (assuming temporal effects are not correlated), or ``fixed'' (no multilevel structure on temporal effects). The contemporaneous effects can also be estimated as being ``fixed'' (all residuals are used to obtain one GGM), ``correlated'' (second step multilevel model with correlated random effects), ``orthogonal'' (second step multilevel model with uncorrelated random effects), or ``unique'' (residuals are used to obtain a GGM per subject). The \emph{mlVAR} package can also be used to plot the estimated networks, in which significance thresholding is used by default with a significance level of $\alpha = 0.05$.

\section{Supplementary 3: Simulation Studies}

In this section, we present simulations to assess the performance of \emph{mlVAR} and  \emph{graphicalVAR} in performing the above-described methods for estimating network structures on ESM data of multiple subjects. Simulation studies on the described methods for cross-sectional and $n=1$ studies are available elsewhere \cite{abegaz2013sparse,qgraphsims,foygel2010extended}. We varied the following conditions:
\begin{itemize}
\item \textbf{Number of nodes.} The number of nodes was set to be either $8$ or $16$, to be representative of plausible variable numbers in psychological ESM studies.
\item \textbf{Sample size.} The sample size was varied between $50$, $100$ and $200$. These values were chosen to represent plausible values in a psychological study.
\item \textbf{Number of observations.} The number of observations per person was varied between $50$, $100$ and $200$. These values were chosen to represent plausible values in a psychological study.
\item \textbf{Fixed structure.} The contemporaneous network structure was simulated to be a chain graph (e.g., $1$ -- $2$ -- $3$) and the between-subjects network was simulated to be a random network with the same number of edges. The temporal network was simulated to be either a chain graph (condition 1) or a chain graph in which every node is connected to the second to next node (e.g., $1 \rightarrow 3 \rightarrow 5$; condition 2). These conditions were chosen such that in condition 1 there is a temporal edge whenever there is a contemporaneous edge, and in condition 2 there is not a temporal edge whenever there is a contemporaneous edge.
\item \textbf{Individual structure.} Edges in the individual networks (temporal and contemporaneous) were either rewired with $50\%$ probability per person or kept stable. The rewiring condition ensured that people may have vastly different network structures, thus reducing the strength of borrowing information from other subjects to estimate individual networks.
\end{itemize}
After generating network structures, and before rewiring, $50\%$ of all edges were made negative. Next, individual edge parameters (for all networks) were drawn from normal distributions with mean of $0.35$ and standard deviation of $0.1$.  Thus, when edges were not rewired, for different subjects in temporal and contemporaneous networks were equal in structure (which edge was present and the sign of the edge) but not in weight. In the condition in which edges were rewired, individual networks differed in both weight and structure. Each condition was replicated $100$ times, leading to $72{,}000$ total simulated datasets. 

We used both \emph{mlVAR} (two-step multi-level VAR) and \emph{graphicalVAR} (pooled and individual LASSO estimation) to estimate fixed and subject-specific network structures. In \emph{mlVAR}, correlated random effects were estimated in the $8$-node condition and orthogonal random effects were estimated in the $16$-node condition. In addition, an ``and''-rule was used to threshold significant edges. In \emph{graphicalVAR}, we varied $10$ (temporal) by $10$ (contemporaneous) LASSO tuning parameters in selecting the optimal GVAR model and $100$ LASSO tuning parameters in selecting the optimal between-subjects GGM. The optimal tuning parameters were selected by minimizing the EBIC with $\gamma = 0.25$. To save computing time, we only estimated one individual subject network per replication in the \emph{graphicalVAR} condition (fixed effects were based on all subjects), and thus base the results of individual network estimation performance in both methods on one network per replication. The true fixed effects were set to the mean of all individual networks created. 

\begin{figure}
\centering
\begin{subfigure}[b]{1\linewidth}
\centering
	  	\includegraphics[width=1\linewidth,page=1]{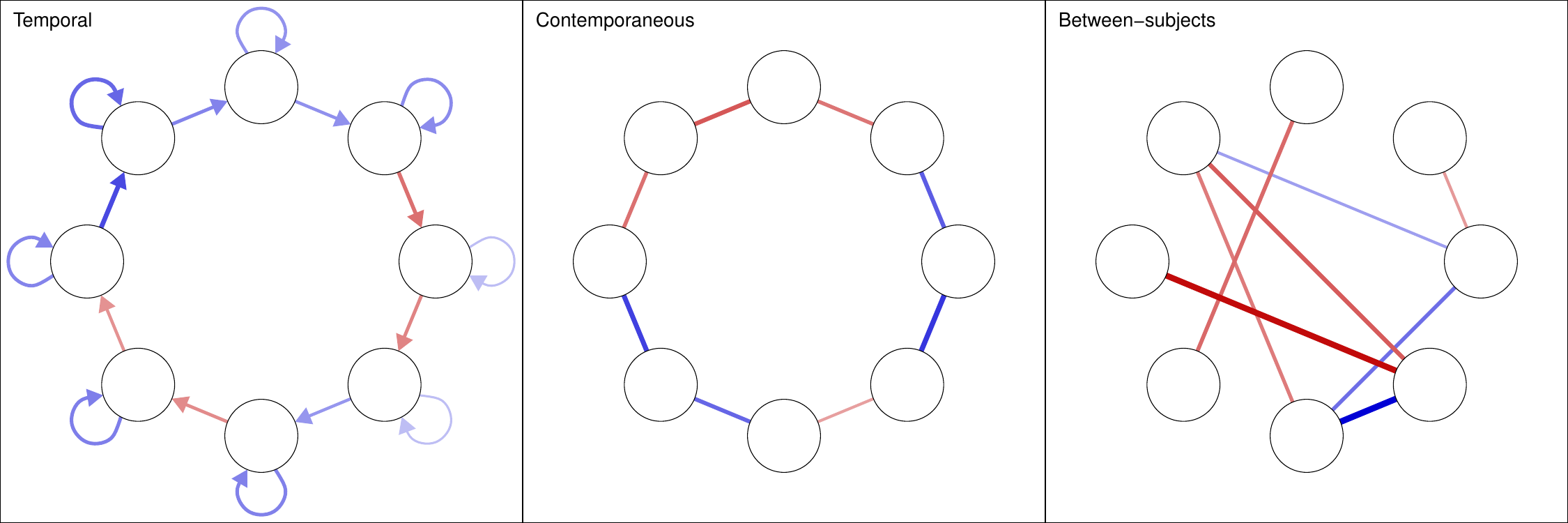}
        \caption{Temporal condition 1 with no re-wiring of within-subject network structures.}
\end{subfigure} \\
\begin{subfigure}[b]{1\linewidth}
\centering
	  	\includegraphics[width=1\linewidth,page=2]{FigureS1.pdf}
        \caption{Temporal condition 1 with $50\%$ re-wiring of within-subject network structures.}
\end{subfigure} \\
\begin{subfigure}[b]{1\linewidth}
\centering
	  	\includegraphics[width=1\linewidth,page=3]{FigureS1.pdf}
        \caption{Temporal condition 2 with no re-wiring of within-subject network structures.}
\end{subfigure}  \\
\begin{subfigure}[b]{1\linewidth}
\centering
	  	\includegraphics[width=1\linewidth,page=4]{FigureS1.pdf}
        \caption{Temporal condition 2 with $50\%$ re-wiring of within-subject network structures.}
\end{subfigure} 
\caption{Examples of generated networks for one subject (temporal and contemporaneous) and overall between-subjects effects. Contemporaneous networks were always simulated to be a chain graph, and temporal networks were simulated to be a chain graph (condition 1) or a chain connected to each second node (e.g., $1 \rightarrow 3 \rightarrow 5$; condition 2). Edges in the within subject network were either rewired (ensuring every person had a different network structure) or kept the same.}
\label{dynamics:fig:simSetup}
\end{figure}

In order to assess how well the estimated networks resemble the true networks, we computed for each dataset the correlations between true and estimated fixed temporal, contemporaneous, and between-subjects networks and the correlations between true and estimated subject-specific temporal and contemporaneous networks---because the between-subjects network does not have random effects. In line with other studies on assessing how well a method retrieves the structure of a network (e.g., \citeNP{epskampPsychometrika,van2014new}), we computed the \emph{sensitivity} (true positive rate) and the \emph{specificity} (true negative rate). In addition, we computed the mean squared error and the average bias (mean absolute deviation of true to estimated parameters) per dataset.

Appendix~A shows the results of the simulation study in the condition where edges were not rewired. It can be seen that performance was generally good in both methods. Fixed effects of the temporal and contemporaneous networks were well estimated (high correlations), most edges in the true network were detected (high sensitivity), and few edges were detected to be nonzero that were, in truth, zero (high specificity). In addition, the bias and mean squared error were generally low. The between-subjects network was better estimated with more people. Using \emph{graphicalVAR} for estimating individual networks showed that at low sample-sizes, the method lacked power to detect true edges (low sensitivity) but did not estimate false edges (high specificity). No model selection is performed in \emph{mlVAR} on subject-specific networks, leading to the specificity of $0$ (all edges were always included in the network). The between-subjects estimation using  \emph{graphicalVAR} featured a moderate specificity, indicating some false edges were detected. It should be noted that the simulations used EBIC tuning parameter $\gamma = 0.25$, which errs more on the side of discovery than the often used $\gamma = 0.5$ value \cite{foygel2010extended}. Of note is that the two-step multi-level procedure performed comparable, if not better, in estimating between-subject network structures than LASSO estimation based on the aggregated scores per person. Appendix~B shows the results in the condition where edges were rewired, and shows here too a good performance for both methods. \emph{mlVAR} estimation performed poorer than when the structure was the same over all subjects, and \emph{graphicalVAR} performed identically. This was expected given that the \emph{graphicalVAR} method does not take information of other subjects into account when estimating a network for one subject, while the  \emph{graphicalVAR} method does. A table with further detailed results from the simulation studies can be found in the other supplementary materials.

\section{Supplementary 4: Stationary distribution}
\label{proofStat}

The GVAR model implies the following expression for the variance-covariance matrix of $\pmb{y}_{t}$ (dropping matrix indexing subscripts for notational clarity):
\[
   \begin{aligned}
    \mathrm{Var}\left(\pmb{y}_T \right) &=  \mathrm{Var}\left(\pmb{B} \pmb{y}_{T-1} + \pmb{\varepsilon}_{T} \right) \\
      \pmb{\Sigma} &= \pmb{B} \pmb{\Sigma} \pmb{B}^{\top} + \pmb{\Theta},
   \end{aligned}
\]
in which we make use of the assumption of stationarity and the assumption that residuals $\pmb{\varepsilon}_{T}$ are uncorrelated with $\pmb{y}_{T-1}$. Now, we can make use of the vectorization operator $\mathrm{Vec}$ and the Kronecker product $\otimes$ to obtain \cite{kim1999state}:
\[
        \begin{aligned}
            \left( \pmb{I} - \pmb{B} \otimes \pmb{B}\right)^{-1} \mathrm{Vec}\left(\pmb{\Theta}\right) &= \mathrm{Vec}\left(\pmb{\Sigma}\right)
         \end{aligned},
\]
which gives an expression for the elements of $\pmb{\Sigma}$ in terms of $\pmb{B}$ and $\pmb{\Theta}$.


\onecolumn
\section{Supplementary Appendix A. Simulation results (no rewiring)}


\begin{center}
\includegraphics[width=0.9\textwidth,page=1]{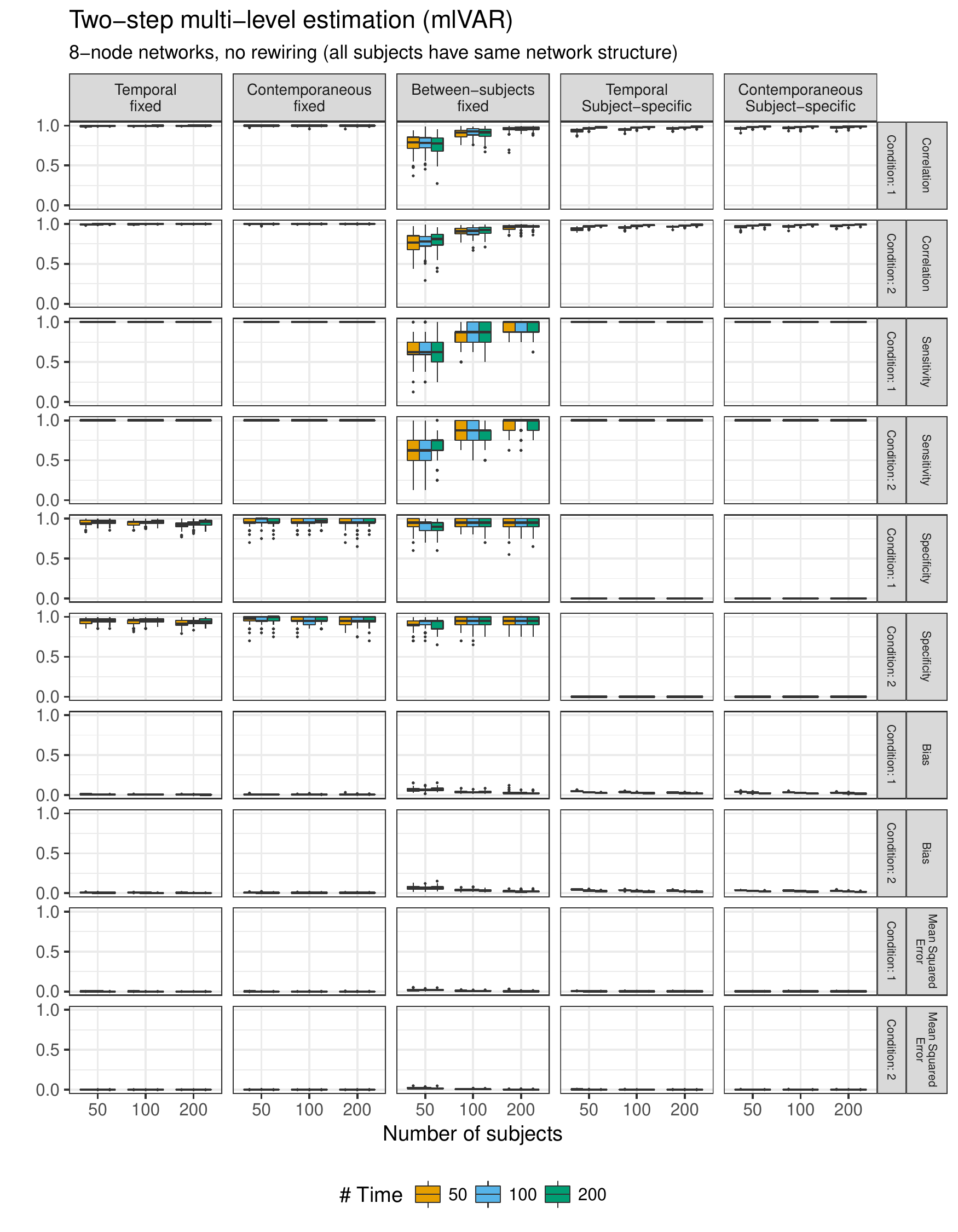}
\end{center}

\begin{center}
\includegraphics[width=0.9\textwidth,page=2]{simStudy_figures1.pdf}
\end{center}

\begin{center}
\includegraphics[width=0.9\textwidth,page=3]{simStudy_figures1.pdf}
\end{center}

\begin{center}
\includegraphics[width=0.9\textwidth,page=4]{simStudy_figures1.pdf}
\end{center}

\section{Supplementary Appendix B. Simulation results ($50\%$ rewiring}


\begin{center}
\includegraphics[width=0.9\textwidth,page=1]{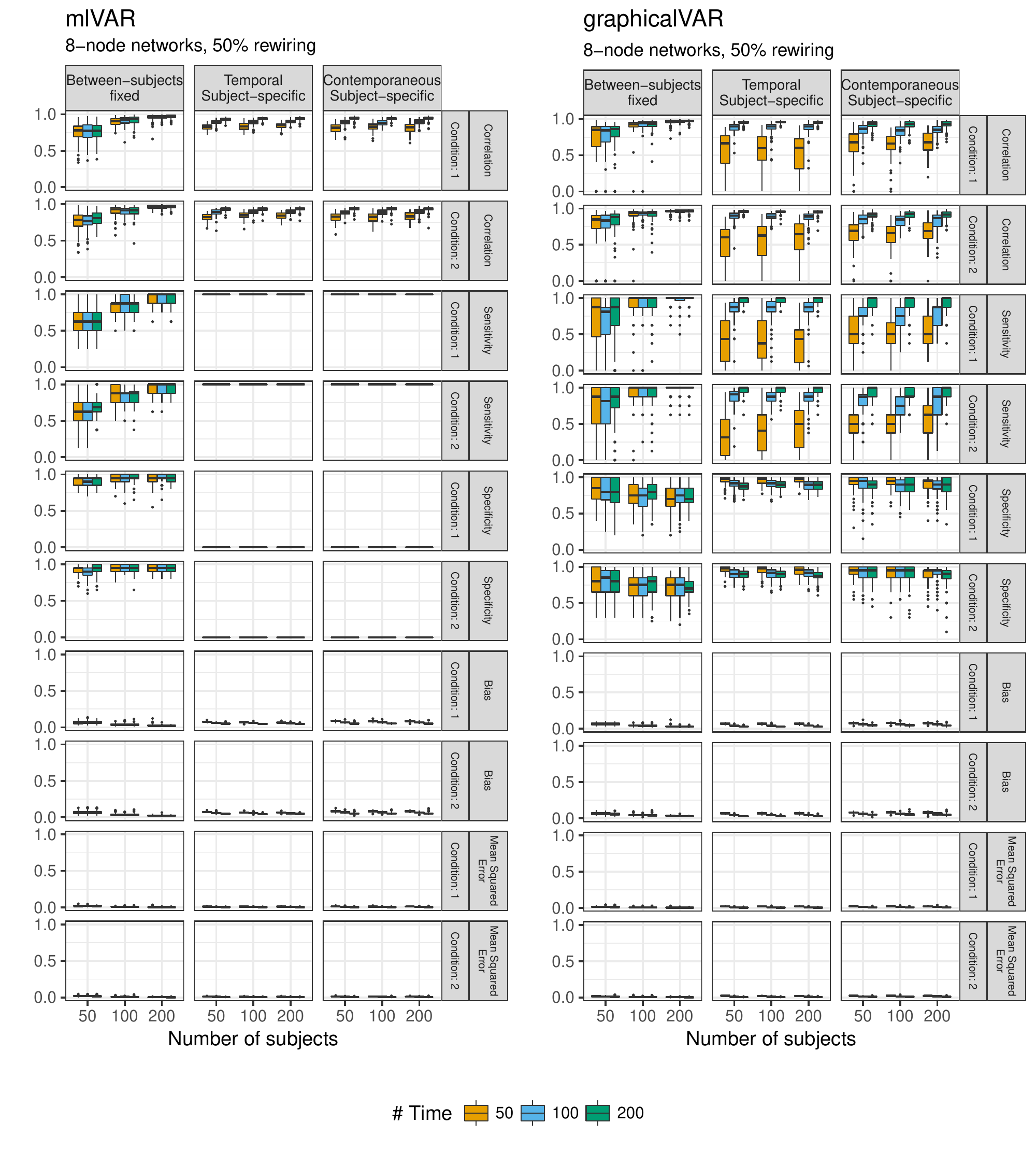}
\end{center}

\begin{center}
\includegraphics[width=0.9\textwidth,page=2]{simStudy_figures2.pdf}
\end{center}

\bibliographystyle{bpacite}
\bibliography{Bibliography}

\end{document}